\documentclass[aps,a4paper,superscriptaddress,showpacs,preprintnumbers,amsmath,amssymb]{revtex4}

\usepackage{ulem}

\usepackage{psfrag} \usepackage{graphicx} \usepackage{dcolumn}
\usepackage{color} \usepackage{latexsym,amsfonts} \usepackage{bm}
\usepackage{amssymb}
\begin{document}

\title{Test of the Standard Model \\ in Neutron Beta Decay with
  Polarized Electron\\ and Unpolarized Neutron and Proton}

\author{A. N. Ivanov}\email{ivanov@kph.tuwien.ac.at}
\affiliation{Atominstitut, Technische Universit\"at Wien, Stadionallee
  2, A-1020 Wien, Austria}
\author{R. H\"ollwieser}\email{roman.hoellwieser@gmail.com}
\affiliation{Atominstitut, Technische Universit\"at Wien, Stadionallee
  2, A-1020 Wien, Austria}\affiliation{Department of Physics,
  Bergische Universit\"at Wuppertal, Gaussstr. 20, D-42119 Wuppertal,
  Germany} \author{N. I. Troitskaya}\email{natroitskaya@yandex.ru}
\affiliation{Atominstitut, Technische Universit\"at Wien, Stadionallee
  2, A-1020 Wien, Austria}
\author{M. Wellenzohn}\email{max.wellenzohn@gmail.com}
\affiliation{Atominstitut, Technische Universit\"at Wien, Stadionallee
  2, A-1020 Wien, Austria} \affiliation{FH Campus Wien, University of
  Applied Sciences, Favoritenstra\ss e 226, 1100 Wien, Austria}
\author{Ya. A. Berdnikov}\email{berdnikov@spbstu.ru}\affiliation{Peter
  the Great St. Petersburg Polytechnic University, Polytechnicheskaya
  29, 195251, Russian Federation}

\date{\today}

\begin{abstract}
We calculate the correlation coefficients of the electron--energy and
electron--antineutrino angular distribution of the neutron
$\beta^-$--decay with polarized electron and unpolarised neutron and
proton.  The calculation is carried out within the Standard Model (SM)
with the contributions, caused by the weak magnetism, proton recoil
and radiative corrections of order of $10^{-3}$, Wilkinson's
corrections of order $10^{-5}$ (Wilkinson, Nucl. Phys. A {\bf 377},
474 (1982) and Ivanov {\it et al.}, Phys. Rev. C {\bf 95}, 055502
(2017)) and the contributions of interactions beyond the SM. The
obtained results can be used for the analysis of experimental data on
searches of interactions beyond the SM at the level of $10^{-4}$
(Abele, Hyperfine Interact. {\bf 237}, 155 (2016)). The contributions
of $G$--odd correlations are calculated and found at the level of
$10^{-5}$ in agreement with the results obtained by Gardner and
Plaster (Phys. Rev. C {\bf 87}, 065504 (2013)) and Ivanov {\it et al.}
(Phys. Rev. C {\bf 98}, 035503 (2018)).
\end{abstract}
\maketitle

\section{Introduction}
\label{sec:introduction}

In Refs.\cite{Ivanov2013,Ivanov2017b,Ivanov2017d} we have calculated
the neutron lifetime and correlation coefficients of the
electron--energy and angular distributions of the neutron
$\beta^-$--decay with polarized neutron and unpolarized electron and
proton, and polarized neutron and electron and unpolarized proton,
respectively. The neutron lifetime and correlation coefficients are
calculated at the level of $10^{-3}$ of contributions of the weak
magnetism and proton recoil of order $O(E_e/M)$, where $E_e$ is the
electron energy and $M$ is an averaged nucleon mass, and radiative
corrections of order $O(\alpha/\pi)$, where $\alpha$ is the
fine--structure constant \cite{PDG2018}. The radiative corrections of
order $O(\alpha/\pi)$ to the neutron lifetime and correlation
coefficients of the neutron $\beta^-$--decay with polarized neutron
and unpolarized electron and proton have been calculated by Sirlin
\cite{Sirlin1967} and Shann \cite{Shann1971} (for details of these
calculations we relegate a reader to \cite{Gudkov2006} and
\cite{Ivanov2013}). In turn, the radiative corrections of order
$O(\alpha/\pi)$ to the correlation coefficients of the neutron
$\beta^-$--decay with polarized neutron and electron, and unpolarized
proton have been calculated in \cite{Ivanov2017b}. Then, in
\cite{Ivanov2013} and \cite{Ivanov2017d} we have taken into account
the contributions of interactions beyond the Standard Model (SM) to
the neutron $\beta^-$--decay with polarized neutron and unpolarized
electron and proton, and polarized neutron and electron, and
unpolarized proton, respectively.

This paper is addressed to the calculation of the correlation
coefficients of the electron--energy and electron--antineutrino
angular distribution of the neutron $\beta^-$--decay with polarized
electron and unpolarized neutron and proton. We calculate a complete
set of corrections of order $10^{-3}$ defined by the corrections of
order $O(E_e/M)$, caused by the weak magnetism and proton recoil and
calculated to next--to--leading order in the large nucleon mass
expansion, and radiative corrections of order $O(\alpha/\pi)$,
calculated to leading order in the large nucleon mass expansion. We
discuss also Wilkinson's corrections of order $10^{-5}$
\cite{Wilkinson1982}, which have been adapted to the neutron
$\beta^-$--decay with polarized neutron and electron and unpolarized
proton in Ref.\cite{Ivanov2017b}. In addition we take into account the
contributions of interactions beyond the SM
\cite{Lee1956}--\cite{Gardner2013} (see also
\cite{Ivanov2013,Ivanov2017d}) including the contributions of the
second class currents (or the $G$--odd correlations)
\cite{Gardner2001,Gardner2013}) (see also \cite{Ivanov2017d}).

The paper is organized as follows. In section \ref{sec:definition} we
write down the general expression for the electron--energy and
electron--antineutrino angular distribution of the neutron
$\beta^-$--decay with polarized electron and unpolarized neutron and
proton. In section \ref{sec:interaction} we discuss the
renormalization procedure of the amplitude of the neutron
$\beta^-$--decay, caused by the effective $V - A$ weak interaction and
radiative corrections, calculated to order $O(\alpha/\pi)$ in the
one--photon exchange approximation. In section \ref{sec:distribution}
we calculate the renormalized electron--energy and
electron--antineutrino angular distribution to order $O(E_e/M)$ and
$O(\alpha/\pi)$, caused by the weak magnetism, proton recoil and
radiative corrections, dependent on the infrared cut--off $\mu$ and
obtained within the finite--photon mass regularization
\cite{Sirlin1967,Ivanov2013}. In section \ref{sec:coulomb} using the
Dirac wave function of the decay electron, distorted in the Coulomb
field of the decay proton, we calculate the correlation coefficient
$L(E_e)$, responsible for time reversal violation.  In section
\ref{sec:spectrum} we write down the observable electron--energy and
electron--antineutrino angular distribution, calculated in the SM to
order $10^{-3}$, caused by the weak magnetism and proton recoil of
order $O(E_e/M)$ and radiative corrections of order
$O(\alpha/\pi)$. We show that the radiative corrections to the
correlation coefficients $H(E_e)$ and $K(E_e)$ are defined by the
functions $(\alpha/\pi)\,h^{(3)}_n(E_e)$ and
$(\alpha/\pi)\,h^{(4)}_n(E_e)$, calculated for the first time in the
present paper. The radiative corrections
$(\alpha/\pi)\,h^{(3)}_n(E_e)$ and $(\alpha/\pi)\,h^{(4)}_n(E_e)$ are
calculated in the Appendix and plotted in Fig.\,\ref{fig:fig3}. In
section \ref{sec:order} we adduce the analytical expressions for the
correlation coefficients $a(E_e)$, $G(E_e)$, $H(E_e)$, $K_e(E_e)$ and
$L(E_e)$, calculated in the SM to order $10^{-3}$, caused by the weak
magnetism, proton recoil and radiative corrections. The obtained
results can be used for the analysis of the experimental data on the
neutron $\beta^-$--decay with polarized electron and unpolarized
neutron and proton. In section \ref{sec:wilkinson} we discuss
Wilkinson's corrections of order $10^{-5}$, which have not been taken
into account for the calculation of the correlation coefficients in
section \ref{sec:order}. They are caused by i) the proton recoil in
the Coulomb electron--proton final--state interaction, ii) the finite
proton radius, iii) the proton--lepton convolution and iv) the
higher--order {\it outer} radiative corrections \cite{Wilkinson1982}.
We calculate the contributions to the correlation coefficients,
induced by the change of the Fermi function caused by the proton
recoil in the electron--proton final--state Coulomb interaction. We
plot these corrections in the electron--energy region $0.761\,{\rm
  MeV} \le E_e \le 0.966\,{\rm MeV}$ in Fig.\,\ref{fig:fig4}. We point
out that Wilkinson's corrections of order $10^{-5}$, caused by ii) the
finite proton radius, iii) the proton--lepton convolution and iv) the
higher--order {\it outer} radiative corrections and calculated in
\cite{Ivanov2017b}, retain fully their shapes and values for the
correlation coefficients analysed in the present paper. In sections
\ref{sec:bsm} and \ref{sec:gparity} we calculate the contributions to
the correlation coefficients, caused by interactions beyond the SM
\cite{Lee1956}--\cite{Gardner2013} (see also
\cite{Ivanov2013,Ivanov2017d}), and give the correlation coefficients
in the form suitable for the analysis of experimental data on searches
of contributions of interactions beyond the SM \cite{Abele2016} (see
also \cite{Ivanov2013,Ivanov2017d}). In section \ref{sec:conclusion}
we discuss the obtained results and perspectives of the theoretical
background to order $10^{-5}$, which goes beyond the scope of
Wilkinson's corrections of order $10^{-5}$
\cite{Ivanov2017b,Ivanov2017c}. In the Appendix we
  calculate the electron--energy and electron--antineutrino angular
  distribution of the neutron radiative $\beta^-$--decay with
  polarized electron and unpolarized neutron and proton. We use these
  results for a cancellation of the infrared divergences in the
  electron--energy and electron--antineutrino angular distribution of
  the neutron $\beta^-$--decay with polarized electron and unpolarized
  neutron and proton. The results, obtained in the Appendix can be
  also used for the experimental analysis of the neutron radiative
  $\beta^-$--decay with polarized electron and unpolarized neutron and
  proton.

\section{Electron--energy and  electron--antineutrino angular 
distribution}
\label{sec:definition}

The electron--energy and electron--antineutrino angular distribution
of the neutron $\beta^-$--decay with polarized electron and
unpolarised neutron and proton can be written in the following form
\cite{Jackson1957,Severijns2006}
\begin{eqnarray}\label{eq:1}
\hspace{-0.3in}\frac{d^5 \lambda_n(E_e,\vec{k}_e,\vec{\xi}_e,
  \vec{k}_{\nu})}{dE_e d\Omega_ed\Omega_{\nu}} &=& (1 +
3\lambda^2)\,\frac{G^2_F|V_{ud}|^2}{32\pi^5} (E_0 - E_e)^2 \sqrt{E^2_e
  - m^2_e}\, E_e\,F(E_e, Z = 1)\,\zeta(E_e)\,\Big\{1 +
a(E_e)\,\frac{\vec{k}_e\cdot \vec{k}_{\nu}}{E_e E_{\nu}}\nonumber\\
\hspace{-0.3in}&+& G(E_e)\,\frac{\vec{\xi}_e\cdot \vec{k}_e}{E_e} +
H(E_e)\,\frac{\vec{\xi}_e \cdot \vec{k}_{\nu}}{E_{\nu}} +
K_e(E_e)\,\frac{(\vec{\xi}_e\cdot \vec{k}_e)( \vec{k}_e\cdot
  \vec{k}_{\nu})}{(E_e + m_e)E_e E_{\nu}}+
L(E_e)\,\frac{\vec{\xi}_e\cdot(\vec{k}_e \times
  \vec{k}_{\nu})}{E_eE_{\nu}} + \ldots\Big\}.
\end{eqnarray}
where $d\Omega_e$ and $d\Omega_{\nu}$ are infinitesimal solid angles
of the electron and antineutrino 3--momenta, $\lambda = - 1.2750(9)$
is the axial coupling \cite{Abele2008} (see also
\cite{Abele2013,Abele2018,Czarnecki2018} and
\cite{Ivanov2013,Ivanov2017b,Ivanov2017d}), $G_F = 1.1664\times
10^{-11}\,{\rm MeV}^{-2}$ is the Fermi weak coupling constant, $V_{ud}
= 0.97417(21)$ is the CKM matrix element \cite{PDG2018}, extracted
from the $0^+ \to 0^+$ transitions, $E_0 = (m^2_n - m^2_p + m^2_e)/2
m_n = 1.2926\,{\rm MeV}$ is the end--point energy of the electron
spectrum, calculated for the neutron $m_n = 939.5654\,{\rm MeV}$,
proton $m_p = 938.2721\,{\rm MeV}$ and electron $m_e = 0.5110\,{\rm
  MeV}$ masses \cite{PDG2018}, $\vec{\xi}_e$ is a unit polarization
vector of the electron, and $F(E_e, Z = 1)$ is the relativistic Fermi
function used in \cite{Ivanov2013,Ivanov2017b,Ivanov2017d} and equal
to \cite{Blatt1952}--\cite{Konopinski1966}
\begin{eqnarray}\label{eq:2}
\hspace{-0.3in}F(E_e, Z = 1 ) = \Big(1 +
\frac{1}{2}\gamma\Big)\,\frac{4(2 r_pm_e\beta)^{2\gamma}}{\Gamma^2(3 +
  2\gamma)}\,\frac{\displaystyle e^{\,\pi
 \alpha/\beta}}{(1 - \beta^2)^{\gamma}}\,\Big|\Gamma\Big(1 + \gamma +
 i\,\frac{\alpha }{\beta}\Big)\Big|^2,
\end{eqnarray}
where $\beta = k_e/E_e = \sqrt{E^2_e - m^2_e}/E_e$ is the electron
velocity, $\gamma = \sqrt{1 - \alpha^2} - 1$, $r_p$ is the electric
radius of the proton.  In the numerical calculations we will use $r_p
= 0.841\,{\rm fm}$ \cite{Pohl2010}.

The function $\zeta(E_e)$ and the correlation coefficients $a(E_e)$
and $G(E_e)$ have been calculated in
\cite{Ivanov2013,Ivanov2017b,Ivanov2017d}.  They are defined by the
contributions of order $10^{-3}$ of the SM interactions, Wilkinson's
corrections of order $10^{-5}$ and interactions beyond the SM (see
\cite{Ivanov2013,Ivanov2017b,Ivanov2017d} and \cite{Ivanov2013a}). In
this paper we calculate the correlation coefficients $H(E_e)$,
$K_e(E_e)$ and $L(E_e)$, where the correlation coefficient $L(E_e)$ is
responsible for violation of invariance under transformation of time
reversal. We calculate i) a complete set of corrections of order
$10^{-3}$, caused by the weak magnetism and proton recoil of order
$O(E_e/M)$ and radiative corrections of order $O(\alpha/\pi)$, ii)
Wilkinson's corrections of order $10^{-5}$ \cite{Wilkinson1982} (see
also \cite{Ivanov2013,Ivanov2017b}), iii) contributions of
interactions beyond the SM \cite{Jackson1957}--\cite{Severijns2006}
(see also \cite{Ivanov2013,Ivanov2017d}) and iv) second class
contributions or $G$--odd correlations \cite{Gardner2001,Gardner2013})
(see also \cite{Ivanov2017d}).

\section{Effective low--energy interactions, defining amplitude of neutron $\beta^-$--decay to  order $10^{-3}$ in the SM}
\label{sec:interaction}

In the SM of electroweak interactions the neutron $\beta^-$--decays,
defined in the one--loop approximation with one--virtual--photon
exchanges, are described by the following interactions
\begin{eqnarray}\label{eq:3}
\hspace{-0.3in}{\cal L}_{\rm int}(x) = {\cal L}_{\rm W}(x) + {\cal
  L}_{\rm em}(x).
\end{eqnarray}
Here ${\cal L}_{\rm W}(x)$ is the effective Lagrangian of low--energy
$V - A$ interactions with a real axial coupling constant $\lambda = -
1.2750(9)$ \cite{Abele2008} (see also \cite{Ivanov2013,Ivanov2017b})
\begin{eqnarray}\label{eq:4}
\hspace{-0.3in}{\cal L}_{\rm W}(x) = -
\frac{G_{0F}}{\sqrt{2}}\,V_{ud}\,\Big\{[\bar{\psi}_{0p}(x)\gamma_{\mu}(1 +
  \lambda \gamma^5)\psi_{0n}(x)] + \frac{\kappa}{2 M}
\partial^{\nu}[\bar{\psi}_{0p}(x)\sigma_{\mu\nu}\psi_{0n}(x)]\Big\}
        [\bar{\psi}_{0e}(x)\gamma^{\mu}(1 - \gamma^5)\psi_{0\nu}(x)],
\end{eqnarray}
where $\psi_{0p}(x)$, $\psi_{0n}(x)$, $\psi_{0e}(x)$ and
$\psi_{0\nu}(x)$ are {\it bare} field operators of the proton,
neutron, electron and antineutrino, respectively, $G_{0F}$ is a {\it
  bare} Fermi weak coupling constant, and $\gamma^{\mu} = (\gamma^0,
\vec{\gamma}\,)$ and $\gamma^5$ are the Dirac matrices
\cite{Itzykson1980}; $\kappa = \kappa_p - \kappa_n = 3.7058$ is the
isovector anomalous magnetic moment of the nucleon, defined by the
anomalous magnetic moments of the proton $\kappa_p = 1.7928$ and the
neutron $\kappa_n = - 1.9130$ and measured in nuclear magneton
\cite{PDG2018}, and $M = (m_n + m_p)/2$ is the average nucleon mass.

For the calculation of the radiative corrections to order
$O(\alpha/\pi)$ the Lagrangian of the electromagnetic interaction
${\cal L}_{\rm em}(x)$ we take in the following form \cite{Ivanov2017c}
\begin{eqnarray}\label{eq:5}
{\cal L}_{\rm em}(x) &=& - \frac{1}{4}\,F^{(0)}_{\mu\nu}(x)F^{(0)\mu\nu}(x) -
\frac{1}{2\xi_0}\,\Big(\partial_{\mu}A^{(0)\mu}(x)\Big)^2 \nonumber\\ && +
\bar{\psi}_{0e}(x)(i\gamma^{\mu}\partial_{\mu} - m_{0e})\psi_{0e}(x) - (-
e_0)\, \bar{\psi}_{0e}(x)\gamma^{\mu}\psi_{0e}(x)A^{(0)}_{\mu}(x)\nonumber\\ &&
+ \bar{\psi}_{0p}(x)(i\gamma^{\mu}\partial_{\mu} - m_{0p})\psi_{0p}(x) - (+ e_0)
\bar{\psi}_{0p}(x)\gamma^{\mu}\psi_{0p}(x) A^{(0)}_{\mu}(x),
\end{eqnarray}
where $F^{(0)}_{\mu\nu}(x) = \partial_{\mu}A^{(0)}_{\nu}(x) -
\partial_{\nu}A^{(0)}_{\mu}(x)$ is the electromagnetic field strength
tensor of the {\it bare} (unrenormalized) electromagnetic field
operator $A^{(0)}_{\mu}(x)$; $\psi_{0e}(x)$ and $\psi_{0p}(x)$ are
{\it bare} operators of the electron and proton fields with {\it bare}
masses $m_{0e}$ and $m_{0p}$, respectively; $- e_0$ and $+ e_0$ are
{\it bare} electric charges of the electron and proton, respectively.
Then, $\xi_0$ is a {\it bare} gauge parameter. After the calculation
of the one--loop corrections of order $O(\alpha/\pi)$ a transition to
the renormalized field operators, masses and electric charges is
defined by the Lagrangian
\begin{eqnarray}\label{eq:6}
{\cal L}_{\rm em}(x) &=& - \frac{1}{4}\,F_{\mu\nu}(x)F^{\mu\nu}(x) -
\frac{1}{2\xi}\,\Big(\partial_{\mu}A^{\mu}(x)\Big)^2\nonumber\\ && +
\bar{\psi}_e(x)(i\gamma^{\mu}\partial_{\mu} - m_{e})\psi_e(x) - (- e)\,
\bar{\psi}_e(x) \gamma^{\mu} \psi_e(x) A_{\mu}(x)\nonumber\\ && +
\bar{\psi}_p(x)(i\gamma^{\mu}\partial_{\mu} - m_p )\psi_p(x) - (+ e)\,
\bar{\psi}_p(x) \gamma^{\mu}\psi_p(x) A_{\mu}(x) + \delta {\cal L}_{\rm em}(x),
\end{eqnarray}
where $A_{\mu}(x)$, $\psi_e(x)$ and $\psi_p(x)$ are the renormalized
operators of the electromagnetic, electron and proton fields,
respectively; $m_e$ and $m_p$ are the renormalized masses of the
electron and proton; $e$ is the renormalized electric charge; and
$\xi$ is the renormalized gauge parameter. The Lagrangian $\delta
{\cal L}_{\rm em}(x)$ contains a complete set of the counterterms
\cite{Weinberg1995},
\begin{eqnarray}\label{eq:7}
\hspace{-0.3in}\delta {\cal L}_{\rm em}(x) &=& - \frac{1}{4}\,(Z_3 -
1)\,F_{\mu\nu}(x)F^{\mu\nu}(x) - \frac{Z_3 -
  1}{Z_{\xi}}\,\frac{1}{2\xi}\,\Big(\partial_{\mu}A^{\mu}(x)\Big)^2\nonumber\\\hspace{-0.3in}
&& + (Z^{(e)}_2 - 1)\,\bar{\psi}_e(x)(i\gamma^{\mu}\partial_{\mu} -
m_{e})\psi_e(x) - (Z^{(e)}_1 - 1)\,(- e)\,\bar{\psi}_e (x)\gamma^{\mu}
\psi_e(x) A_{\mu}(x) - Z^{(e)}_2 \delta m_e \bar{\psi}_e(x)\psi_e(x)
\nonumber\\ \hspace{-0.3in}&& + (Z^{(p)}_2 -
1)\,\bar{\psi}_p(x)(i\gamma^{\mu}\partial_{\mu} - m_p )\psi_p(x) -
(Z^{(p)}_1 - 1)\,( + e) \,\bar{\psi}_p(x) \gamma^{\mu}\psi_p(x)
A_{\mu}(x) - Z^{(p)}_2 \delta m_p \bar{\psi}_p(x) \psi_p(x),
\end{eqnarray}
where $Z_3$, $Z^{(e)}_2$, $Z^{(e)}_1$, $Z^{(p)}_2$, $Z^{(p)}_1$,
$\delta m_e$ and $\delta m_p$ are the counterterms. Here $Z_3$ is the
renormalization constant of the electromagnetic field operator
$A_{\mu}$, $Z^{(e)}_2$ and $Z^{(e)}_1$ are the renormalization
constants of the electron field operator $\psi_e$ and the
electron--electron--photon ($e^-e^-\gamma$) vertex, respectively;
$Z^{(p)}_2$ and $Z^{(p)}_1$ are the renormalization constants of the
proton field operator $\psi_p$ and the proton--proton--photon ($p p
\gamma$) vertex, respectively. Then, $(- e)$ and $(+ e)$, $m_e$ and
$m_p$ and $\delta m_e$ and $\delta m_p$ are the renormalized electric
charges and masses and the mass--counterterms of the electron and
proton, respectively. Rescaling the field operators
\cite{Weinberg1995,Bogoliubov1959}
\begin{eqnarray}\label{eq:8}
\sqrt{Z_3}\, A_{\mu}(x) = A^{(0)}_{\mu}(x)\quad,\quad
\sqrt{Z^{(e)}_2}\,\psi_e(x) = \psi_{0e}(x)\quad,\quad
\sqrt{Z^{(p)}_2}\,\psi_p(x) = \psi_{0p}(x)
\end{eqnarray}
and denoting $m_e + \delta m_e = m_{0e}$, $m_p + \delta m_p = m_{0p}$
and $Z_{\xi} \xi = \xi_0$ we arrive at the Lagrangian
\begin{eqnarray}\label{eq:9}
\hspace{-0.3in}{\cal L}_{\rm em}(x) &=& -
\frac{1}{4}\,F^{(0)}_{\mu\nu}(x)F^{(0)\mu\nu}(x) -
\frac{1}{2\xi_0}\,\Big(\partial_{\mu}A^{(0)\mu}(x)\Big)^2\nonumber\\ 
\hspace{-0.3in}&&+
  \bar{\psi}_{0e}(x)(i\gamma^{\mu}\partial_{\mu} - m_{0e})\psi_{0e}(x)
  - ( - e)\,Z^{(e)}_1 (Z^{(e)}_2)^{-1} Z^{-1/2}_3
  \bar{\psi}_{0e}(x)\gamma^{\mu}\psi_{0e}(x)A^{(0)}_{\mu}(x)\nonumber\\ 
\hspace{-0.3in}&& + \bar{\psi}_{0p}(x)(i\gamma^{\mu}\partial_{\mu} -
m_{0p})\psi_{0p}(x) - (+ e)\, Z^{(p)}_1(Z^{(p)}_2)^{-1} Z^{-1/2}_3
\bar{\psi}_{0p}(x)\gamma^{\mu}\psi_{0p}(x)A^{(0)}_{\mu}(x).
\end{eqnarray}
Because of the Ward identities $Z^{(e)}_1 = Z^{(e)}_2$ and $Z^{(p)}_1
= Z^{(p)}_2$ \cite{Itzykson1980,Weinberg1995,Bogoliubov1959}, we may
replace $(-e)\,Z^{-1/2}_3 = - e_0$ and $(+ e)\,Z^{-1/2}_3 = +
e_0$. This brings Eq.(\ref{eq:9}) to the form of Eq.(\ref{eq:5}). We
would like to emphasize that to order $O(\alpha/\pi)$ the
renormalization constant $Z_3$ is equal to unity because of the absent
of closed fermion loops
\cite{Itzykson1980,Weinberg1995,Bogoliubov1959}, i.e., $Z_3 = 1$.
This means that in such an approximation the {\it bare} electric
charge $e_0$ coincides with the renormalized electric charge $e$,
i.e. $e_0 = e$. After the rescaling of the proton and electron field
operators Eq.(\ref{eq:8}) the Lagrangian of $V - A$ weak interactions
Eq.(\ref{eq:4}) takes the form
\begin{eqnarray}\label{eq:10}
\hspace{-0.3in}{\cal L}_{\rm W}(x) = -
\frac{G_F}{\sqrt{2}}\,V_{ud}\,\Big\{[\bar{\psi}_p(x)\gamma_{\mu}(1 +
  \lambda \gamma^5)\psi_n(x)] + \frac{\kappa}{2 M}
\partial^{\nu}[\bar{\psi}_p(x)\sigma_{\mu\nu}\psi_n(x)]\Big\}
        [\bar{\psi}_e(x)\gamma^{\mu}(1 - \gamma^5)\psi_{\nu}(x)],
\end{eqnarray}
where $G_F = \sqrt{Z^{(p)}_2 Z^{(e)}_2}G_{0F}$ is the Fermi weak
coupling constant renormalized by electromagnetic interactions to
order $O(\alpha/\pi)$. The {\it bare} neutron $\psi_{0n}(x)$ and
antineutrino $\psi_{0\nu}(x)$ field operators are not renormalized by
electromagnetic interactions and coincide with the field operators
$\psi_n(x)$ and $\psi_{\nu}(x)$, respectively, i.e.  $\psi_{0n}(x) =
\psi_n(x)$ and $\psi_{0\nu}(x) = \psi_{\nu}(x)$. 

\begin{figure}
\centering \includegraphics[height=0.12\textheight]{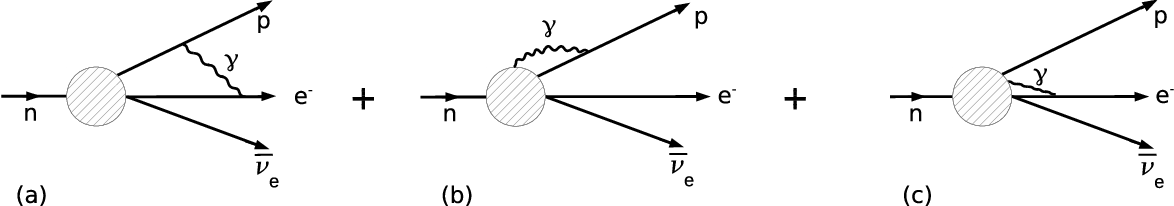}
  \caption{The Feynman diagrams, defining the main contribution of the
    radiative corrections of order $O(\alpha/\pi)$, caused by
    one--virtual photon exchanges, to the neutron $\beta^-$--decay
    (see Sirlin \cite{Sirlin1967}).}
\label{fig:fig1}
\end{figure}
\begin{figure}
\centering \includegraphics[height=0.12\textheight]{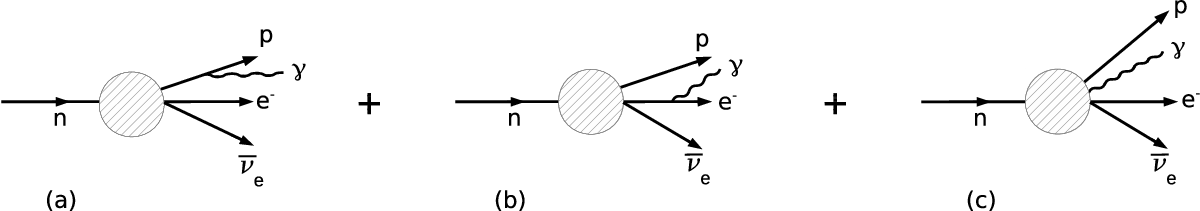}
  \caption{The Feynman diagrams, defining the contribution to the
    amplitude of the neutron radiative $\beta^-$--decay in the
    tree--approximation to order $e$.}
\label{fig:fig2}
\end{figure}

\section{Electron--energy and electron--antineutrino angular 
distribution with radiative corrections caused by one--virtual photon
exchanges}
\label{sec:distribution}

Using the results, obtained in \cite{Ivanov2013}, the renormalized
amplitude of the neutron $\beta^-$--decay with contributions, caused
by the weak magnetism and proton recoil, calculated to
next--to--leading order $O(E_e/M)$ in the large nucleon mass
expansion, and radiative corrections to order $O(\alpha/\pi)$, defined
by the Feynman diagrams in Fig.\,\ref{fig:fig1} and calculated to
leading order in the large nucleon mass expansion, takes the form (see
Eq.(D-52) of Ref.\cite{Ivanov2013})
\begin{eqnarray}\label{eq:11}
\hspace{-0.3in}&&M(n \to p\,e^- \,\bar{\nu}_e) = - 2m_n\,\frac{G_F}
       {\sqrt{2}}V_{ud} \Big\{\Big(1 +
       \frac{\alpha}{2\pi}\,f_{\beta^-_c}(E_e,\mu)\Big)[\varphi^{\dagger}_p
         \varphi_n][\bar{u}_e\,\gamma^0(1 -
         \gamma^5)v_{\bar{\nu}}]\nonumber\\
\hspace{-0.3in}&& - \tilde{\lambda} \Big(1 +
\frac{\alpha}{2\pi}\,f_{\beta^-_c}(E_e,\mu)\Big)[\varphi^{\dagger}_p
  \vec{\sigma}\,\varphi_n]\cdot [\bar{u}_e \vec{\gamma}\,(1 -
  \gamma^5)v_{\bar{\nu}}] -
\frac{\alpha}{2\pi}\,g_F(E_e)\,[\varphi^{\dagger}_p
  \varphi_n][\bar{u}_e\,(1 -
  \gamma^5)v_{\bar{\nu}}]\nonumber\\\hspace{-0.3in}&&+
\frac{\alpha}{2\pi}\,\tilde{\lambda} g_F(E_e) [\varphi^{\dagger}_p
  \vec{\sigma}\,\varphi_n]\cdot [\bar{u}_e \gamma^0\vec{\gamma}\,(1 -
  \gamma^5)v_{\bar{\nu}}] - \frac{m_e}{2
M}\,[\varphi^{\dagger}_p\varphi_n][\bar{u}_e\,(1 -
\gamma^5)v_{\bar{\nu}}]\nonumber\\
\hspace{-0.3in}&& + \frac{\tilde{\lambda}}{2
  M}[\varphi^{\dagger}_p(\vec{\sigma}\cdot \vec{k}_p) \varphi_n
]\,[\bar{u}_e\,\gamma^0 (1 - \gamma^5)v_{\bar{\nu}}] - i\,
\frac{\kappa + 1}{2 M} [\varphi^{\dagger}_p (\vec{\sigma}\times
  \vec{k}_p) \varphi_n] \cdot [\bar{u}_e\,\vec{\gamma}\,(1 -
  \gamma^5)v_{\bar{\nu}}] \Big\},
\end{eqnarray}
where $\varphi_p$ and $\varphi_n$ are Pauli spinorial wave functions
of the proton and neutron, $u_e$ and $v_{\nu}$ are Dirac wave
functions of the electron and electron antineutrino, $\vec{\sigma}$
are the Pauli $2\times 2$ matrices, and $\tilde{\lambda} = \lambda (1
- E_0/2M)$ and $\vec{k}_p = - \vec{k}_e - \vec{k}_{\nu}$ is the proton
3--momentum in the rest frame of the neutron. The functions
$f_{\beta^-_c}(E_e,\mu)$ and $g_F(E_e)$ are equal to (see Eq.(D-51))
\begin{eqnarray}\label{eq:12}
f_{\beta^-_c}(E_e,\mu) &=& \frac{3}{2}{\ell
  n}\Big(\frac{m_p}{m_e}\Big) - \frac{11}{8} + 2 {\ell
  n}\Big(\frac{\mu}{ m_e}\Big)\,\Big[\frac{1}{2\beta}\,{\ell
    n}\Big(\frac{1 + \beta}{1 - \beta}\Big) - 1 \Big] -
\frac{1}{\beta}{\rm Li}_2\Big(\frac{2\beta}{1 + \beta}\Big) -
\frac{1}{4\beta}\,{\ell n}^2\Big(\frac{1 + \beta}{1 -
  \beta}\Big)\nonumber\\ &+& \frac{1}{2\beta}\,{\ell n}\Big(\frac{1 +
  \beta}{1 - \beta}\Big) + C_{WZ},\nonumber\\ g_F(E_e) &=&\frac{\sqrt{1
    - \beta^2}}{2\beta}\,{\ell n}\Big(\frac{1 + \beta}{1 -
  \beta}\Big),
\end{eqnarray}
where $\mu$ is a photon mass, which should be taken in the limit $\mu
\to 0$, and ${\rm Li}_2(x)$ is the Polylogarithmic function. A photon
mass $\mu$ is used for Lorentz invariant regularization of infrared
divergences of radiative corrections \cite{Sirlin1967}. The constant
$C_{WZ}$, defined by the contributions of the $W$--boson and
$Z$--boson exchanges and the QCD corrections \cite{Sirlin2004} (see
also \cite{Sirlin1986,Sirlin2006}), is equal to $C_{WZ} = 10.249$ (see
also discussion below Eq.(D-58) of Ref.\cite{Ivanov2013}).

The squared absolute value of the matrix element Eq.(\ref{eq:12}),
summed over polarizations of massive fermions, we calculate for
polarized electron and unpolarized neutron and proton
\cite{Ivanov2017b}. We get (see also Eq.(A-16) in Appendix A of
Ref.\cite{Ivanov2013})
\begin{eqnarray}\label{eq:13}
\hspace{-0.3in}&&\sum_{\rm pol}\frac{|M(n\to p e^-\nu_e)|^2}{8m^2_n
  G^2_F |V_{ud}|^2} = \Big(1 +
\frac{\alpha}{\pi}\,f_{\beta^-_c}(E_e,\mu)\Big){\rm tr}\{(\hat{k}_e +
m_e)(1 + \gamma^5 \hat{\zeta}_e) \gamma^0 \hat{k}_{\nu} \gamma^0 (1 -
\gamma^5)
\}\nonumber\\ 
\hspace{-0.3in}&&-\frac{\alpha}{2\pi}\,g_F(E_e)\,{\rm
  tr}\{(\hat{k}_e + m_e)(1 + \gamma^5 \hat{\zeta}_e) \gamma^0
\hat{k}_{\nu}(1 + \gamma^5) \} - \frac{\alpha}{2\pi}\,g_F(E_e)\,{\rm
  tr}\{(\hat{k}_e + m_e)(1 + \gamma^5 \hat{\zeta}_e) \hat{k}_{\nu}
\gamma^0(1 - \gamma^5) \}\nonumber\\
\hspace{-0.3in}&&+ \tilde{\lambda}^2\Big(1 +
\frac{\alpha}{\pi}\,f_{\beta^-_c}(E_e,\mu)\Big)\delta^{ij}{\rm tr}\{
(\hat{k}_e + m_e)(1 + \gamma^5 \hat{\zeta}_e)
\gamma^j\hat{k}_{\nu}\gamma^i (1 - \gamma^5)\}\nonumber\\
\hspace{-0.3in}&&-
\tilde{\lambda}^2\,\frac{\alpha}{2\pi}\,g_F(E_e)\,\delta^{ij}{\rm
  tr}\{(\hat{k}_e + m_e)(1 + \gamma^5 \hat{\zeta}_e) \gamma^0 \gamma^i
\hat{k}_{\nu}\gamma^j (1 - \gamma^5) \}\nonumber\\
\hspace{-0.3in}&& + \tilde{\lambda}^2\,
\frac{\alpha}{2\pi}\,g_F(E_e)\,\delta^{ij}{\rm tr}\{(\hat{k}_e +
m_e)(1 + \gamma^5 \hat{\zeta}_e)\gamma^i \hat{k}_{\nu}
\gamma^0\gamma^j (1 +\gamma^5) \}\nonumber\\
\hspace{-0.3in}&&- \frac{m_e}{2M}\,{\rm tr}\{(\hat{k}_e + m_e) (1 +
\gamma^5 \hat{\zeta}_e)\gamma^0 \hat{k}_{\nu} (1 + \gamma^5)\} -
\frac{m_e}{2M}{\rm tr}\{(\hat{k}_e + m_e)(1 + \gamma^5 \hat{\zeta}_e)
\hat{k}_{\nu} \gamma^0(1 - \gamma^5)\}\nonumber\\
\hspace{-0.3in}&&- \frac{\tilde{\lambda}^2}{2 M}\,\vec{k}_p\cdot {\rm
  tr}\{(\hat{k}_e + m_e)(1 + \gamma^5 \hat{\zeta}_e) \vec{\gamma}\,
\hat{k}_{\nu} \gamma^0 (1 - \gamma^5)\} -
\frac{\tilde{\lambda}^2}{2 M}\,\vec{k}_p\cdot {\rm tr}\{(\hat{k}_e +
m_e)(1 + \gamma^5 \hat{\zeta}_e)\gamma^0 \hat{k}_{\nu} \,\vec{\gamma}\,(1 -
\gamma^5)\}\nonumber\\
\hspace{-0.3in}&&+ i\,\tilde{\lambda} \frac{\kappa +
  1}{2M}\,\varepsilon^{\ell j a}k^a_p\,{\rm tr}\{(\hat{k}_e + m_e)(1 +
\gamma^5 \hat{\zeta}_e) \gamma^{\ell}\,\hat{k}_{\nu}\gamma^j (1 -
\gamma^5)\}\nonumber\\
\hspace{-0.3in}&& - i\,\tilde{\lambda} \frac{\kappa +
  1}{2M}\,\varepsilon^{\ell j a}k^a_p\,{\rm tr}\{(\hat{k}_e + m_e)(1 +
\gamma^5 \hat{\zeta}_e) \gamma^j \hat{k}_{\nu} \gamma^{\ell} (1 -
\gamma^5)\},
\end{eqnarray}
where $\zeta^{\mu}_e = (\zeta^0_e, \vec{\zeta}_e)$ is the 4--vector of
an electron polarization defined by \cite{Ivanov2017b}
\begin{eqnarray}\label{eq:14}
\zeta^{\mu}_e = (\zeta^0_e, \vec{\zeta}_e) = \Big(\frac{\vec{k}_e\cdot
  \vec{\xi}_e}{m_e}, \vec{\xi}_e + \frac{\vec{k}_e(\vec{k}_e\cdot
  \vec{\xi}_e)}{m_e(E_e + m_e)}\Big).
\end{eqnarray}
It obeys the constraints $\zeta^2_e = -1$ and $k_e\cdot \zeta_e = 0$,
where $\vec{\xi}_e$ is a unit vector of the electron polarization
\cite{Itzykson1980}. We would like to emphasize that in
Eq.(\ref{eq:13}) following Sirlin \cite{Sirlin1967} we have neglected
the contributions of order $O(\alpha E_e/\pi M)$. Having calculated
the traces over Dirac matrices we obtain
\begin{eqnarray}\label{eq:15}
\hspace{-0.3in}&&\sum_{\rm pol}\frac{|M(n\to p e^-\nu_e)|^2}{32m^2_n
  G^2_F |V_{ud}|^2E_e E_{\nu}} = \Big(1 +
\frac{\alpha}{\pi}\,f_{\beta^-_c}(E_e,\mu)\Big)\Big(1 +
\frac{\vec{k}_e\cdot \vec{k}_{\nu}}{E_e E_{\nu}} -
\frac{\vec{\xi}_e\cdot \vec{k}_e}{E_e} -
\frac{m_e}{E_e}\,\frac{\vec{\xi}_e\cdot \vec{k}_{\nu}}{E_{\nu}} -
\frac{(\vec{\xi}_e\cdot \vec{k}_e)(\vec{k}_e\cdot \vec{k}_{\nu})}{(E_e
  + m_e)E_eE_{\nu}}\Big)\nonumber\\
\hspace{-0.3in}&& -\frac{\alpha}{\pi}\,g_F(E_e) \Big(\frac{m_e}{E_e} +
\frac{\vec{k}_e\cdot \vec{k}_{\nu}}{E_e E_{\nu}} -
\frac{\vec{\xi}_e\cdot \vec{k}_{\nu}}{E_{\nu}}-
\frac{E_e}{m_e}\,\frac{(\vec{\xi}_e\cdot \vec{k}_e)(\vec{k}_e\cdot
  \vec{k}_{\nu})}{(E_e + m_e)E_eE_{\nu}}\Big)\nonumber\\
\hspace{-0.3in}&&+ \tilde{\lambda}^2\Big(1 +
\frac{\alpha}{\pi}\,f_{\beta^-_c}(E_e,\mu)\Big)\Big(3 -
\frac{\vec{k}_e\cdot \vec{k}_{\nu}}{E_e E_{\nu}} -
3\frac{\vec{\xi}_e\cdot \vec{k}_e}{E_e} +
\frac{m_e}{E_e}\frac{\vec{\xi}_e\cdot \vec{k}_{\nu}}{E_{\nu}} +
\frac{(\vec{\xi}_e\cdot \vec{k}_e)(\vec{k}_e\cdot \vec{k}_{\nu})}{(E_e
  + m_e)E_eE_{\nu}}\Big)\nonumber\\
\hspace{-0.3in}&&-
\tilde{\lambda}^2\,\frac{\alpha}{\pi}\,g_F(E_e)\,\Big(3\frac{m_e}{E_e}
+ \frac{\vec{\xi}_e\cdot \vec{k}_{\nu}}{E_{\nu}} -
\frac{(\vec{\xi}_e\cdot \vec{k}_e)(\vec{k}_e\cdot \vec{k}_{\nu})}{(E_e
  + m_e)E_eE_{\nu}}\Big) - \frac{m_e}{M}\,\Big(\frac{m_e}{E_e} -
\frac{\vec{\xi}_e\cdot \vec{k}_{\nu}}{E_{\nu}} +
\frac{(\vec{\xi}_e\cdot \vec{k}_e)(\vec{k}_e\cdot \vec{k}_{\nu})}{(E_e
  + m_e)E_eE_{\nu}}\Big)\nonumber\\
\hspace{-0.3in}&&+ \frac{\tilde{\lambda}^2}{M}\,\Big(\Big(E_0 -
\frac{m^2_e}{E_e}\Big) + E_0\,\frac{\vec{k}_e\cdot \vec{k}_{\nu}}{E_e
  E_{\nu}} - E_0\,\frac{\vec{\xi}_e\cdot \vec{k}_e}{E_e} -
E_{\nu}\,\frac{m_e}{E_e}\,\frac{\vec{\xi}_e\cdot
  \vec{k}_{\nu}}{E_{\nu}} - (E_0 + m_e)\frac{(\vec{\xi}_e\cdot
  \vec{k}_e)(\vec{k}_e\cdot \vec{k}_{\nu})}{(E_e +
  m_e)E_eE_{\nu}}\Big)\nonumber\\
\hspace{-0.3in}&&+ \tilde{\lambda} \frac{2(\kappa +
  1)}{M}\,\Big(\Big(E_0 - 2 E_e + \frac{m^2_e}{E_e}\Big) + (2 E_e -
E_0)\,\frac{\vec{k}_e\cdot \vec{k}_{\nu}}{E_e E_{\nu}} + (2 E_e -
E_0)\,\frac{\vec{\xi}_e\cdot \vec{k}_e}{E_e} + (E_0 -
E_e)\,\frac{m_e}{E_e}\,\frac{\vec{\xi}_e\cdot \vec{k}_{\nu}}{E_{\nu}}
\nonumber\\
\hspace{-0.3in}&& + (E_0 - m_e)\,\frac{(\vec{\xi}_e\cdot
  \vec{k}_e)(\vec{k}_e\cdot \vec{k}_{\nu})}{(E_e +
  m_e)E_eE_{\nu}}\Big),
\end{eqnarray}
where we have used a relation $E_e + E_{\nu} = E_0$. Now we have to
take into account the contribution of the phase--volume
\cite{Ivanov2013} and multiply Eq.(\ref{eq:15}) by the function
\begin{eqnarray}\label{eq:16}
\Phi_{\beta^-_c}(\vec{k}_e,\vec{k}_{\nu}) = 1 + \frac{3}{M}\Big(E_e -
\frac{\vec{k}_e\cdot \vec{k}_{\nu}}{E_{\nu}}\Big).
\end{eqnarray}
This gives
\begin{eqnarray}\label{eq:17}
\hspace{-0.3in}&&\Phi_{\beta^-_c}(\vec{k}_e,\vec{k}_{\nu})\sum_{\rm
  pol}\frac{|M(n\to p e^-\nu_e)|^2}{32m^2_n G^2_F |V_{ud}|^2 E_e
  E_{\nu}} = (1 + 3\lambda^2)\tilde{\zeta}(E_e)\Big\{1 +
\tilde{a}(E_e)\frac{\vec{k}_e\cdot \vec{k}_{\nu}}{E_e E_{\nu}} +
\tilde{G}(E_e)\,\frac{\vec{\xi}_e\cdot \vec{k}_e}{E_e} +
\tilde{H}(E_e)\,\frac{\vec{\xi}_e \cdot
  \vec{k}_{\nu}}{E_{\nu}}\nonumber\\
\hspace{-0.3in}&&+ \tilde{K}_e(E_e)\,\frac{(\vec{\xi}_e\cdot
  \vec{k}_e)( \vec{k}_e\cdot \vec{k}_{\nu})}{(E_e + m_e)E_e E_{\nu}} -
3\,a_0\, \frac{E_e}{M}\,\Big(\frac{(\vec{k}_e\cdot
  \vec{k}_{\nu})^2}{E^2_e E^2_{\nu}} -
\frac{1}{3}\,\frac{k^2_e}{E^2_e}\,\Big) + 3\,a_0\, 
\frac{m_e}{M}\,\Big(\frac{(\vec{\xi}_e\cdot
  \vec{k}_{\nu})(\vec{k}_e\cdot \vec{k}_{\nu})}{E_e E^2_{\nu}} -
\frac{1}{3}\,\frac{\vec{\xi}_e\cdot \vec{k}_e}{E_e}\,\Big)\nonumber\\
\hspace{-0.3in}&&+ 3\,a_0\,
\frac{1}{M}\,\Big(\frac{(\vec{\xi}_e\cdot \vec{k}_e)(\vec{k}_e\cdot
  \vec{k}_{\nu})^2}{(E_e + m_e)E_e E^2_{\nu}} - \frac{1}{3}\,(E_e -
m_e)\,\frac{\vec{\xi}_e\cdot \vec{k}_e}{E_e}\,\Big) \Big\},
\end{eqnarray}
where we have denoted $a_0 = (1 - \lambda^2)/(1 + 3\lambda^2)$ and
\begin{eqnarray*}
\hspace{-0.3in}\tilde{\zeta}(E_e) &=& \Big(1 +
\frac{\alpha}{\pi}\,f_{\beta^-_c}(E_e,\mu) -
\frac{\alpha}{\pi}\,g_F(E_e)\,\frac{m_e}{E_e}\Big) +
\frac{1}{M}\,\frac{1}{1 + 3\lambda^2}\,\Big[- 2\,\lambda\Big(\lambda -
  (\kappa + 1)\Big)\,E_0\nonumber\\
\hspace{-0.3in}&&+ \Big(10 \lambda^2 - 4(\kappa + 1)\, \lambda
  + 2\Big)\,E_e- 2 \lambda\,\Big(\lambda - (\kappa +
1)\Big)\,\frac{m^2_e}{E_e}\Big],\nonumber\\
\hspace{-0.3in}\tilde{\zeta}(E_e)\,\tilde{a}(E_e) &=& a_0 \Big(1 +
\frac{\alpha}{\pi}\,f_{\beta^-_c}(E_e,\mu)\Big) +
\frac{1}{M}\,\frac{1}{1 + 3\lambda^2}\,\Big[2 \lambda \Big(\lambda -
  (\kappa + 1)\Big)E_0 - 4\lambda \Big(3\lambda - (\kappa + 1)\Big)
E_e\Big],\nonumber\\
\hspace{-0.3in}\tilde{\zeta}(E_e)\,\tilde{G}(E_e) &=& - \Big(1 +
\frac{\alpha}{\pi}\,f_{\beta^-_c}(E_e,\mu) \Big) +
\frac{1}{M}\,\frac{1}{1 + 3\lambda^2}\,\Big[2\lambda \Big(\lambda -
  (\kappa + 1)\Big)\,E_0 - \Big(10 \lambda^2 - 4(\kappa + 1)\,\lambda
  + 2\Big)E_e\Big]\nonumber\\
\hspace{-0.3in}\tilde{\zeta}(E_e)\tilde{H}(E_e) &=&
\frac{m_e}{E_e}\,\Big\{- a_0 \Big(1 +
\frac{\alpha}{\pi}\,f_{\beta^-_c}(E_e,\mu) -
\frac{\alpha}{\pi}\,g_F(E_e)\,\frac{E_e}{m_e}\Big) +
\frac{1}{M}\,\frac{1}{1 + 3\lambda^2}\,\Big[ - 2\lambda\Big(\lambda -
  (\kappa + 1)\Big) E_0\nonumber\\
\hspace{-0.3in}&& + \Big(4\lambda^2 - 2 (\kappa + 1)\lambda - 2\Big)
E_e\Big]\Big\},\nonumber\\
\hspace{-0.3in}\tilde{\zeta}(E_e)\tilde{K}_e(E_e) &=& - a_0 \Big(1 +
\frac{\alpha}{\pi}\,f_{\beta^-_c}(E_e,\mu) +
\frac{\alpha}{\pi}\,g_F(E_e)\Big) +
\frac{1}{M}\,\frac{1}{1 + 3\lambda^2}\,\Big[- 2\lambda\Big(\lambda -
  (\kappa + 1)\Big) E_0 \nonumber\\
\end{eqnarray*}
\begin{eqnarray}\label{eq:18}
\hspace{-0.3in}&&+ 4\lambda\Big(3\lambda - (\kappa + 1)\Big)E_e +
\Big(8\lambda^2 - 2 (\kappa + 1)\lambda + 2\Big) m_e\Big].
\end{eqnarray}
The use of the Dirac wave function of a free decay electron leads to a
vanishing correlation coefficient $\tilde{L}(E_e) = 0$. In order to
get a non--vanishing correlation coefficient $\tilde{L}(E_e)$ we have
to use the Dirac wave function of a decay electron, distorted in the
Coulomb field of the decay proton
\cite{Jackson1958,Konopinski1966,Ivanov2014}. 

\section{Correlation coefficient $L(E_e)$}
\label{sec:coulomb}

For the calculation of the correlation coefficient we use the Dirac wave
function of the electron, distorted by the Coulomb proton--electron
final state interaction. It is equal to
\cite{Jackson1958,Konopinski1966,Ivanov2014}
\begin{eqnarray}\label{eq:19}
\hspace{-0.3in}u_e(\vec{k}_e, \sigma_e) = \sqrt{\frac{E_e + m_e(1 -
    \gamma)}{1 - \gamma}}\left(\begin{array}{c}1 \\ \displaystyle
  \Big(1 + i\,\frac{\alpha Z m_e}{k_e}\Big)\frac{\vec{\sigma}\cdot
    \vec{k}_e}{E_e + m_e(1 - \gamma)}
\end{array}\right)\otimes \varphi_{\sigma_e},
\end{eqnarray}
where $\gamma = 1 - \sqrt{1 - \alpha^2 Z^2}$. The electron wave
function Eq.(\ref{eq:19}) satisfies the Dirac equation
\cite{Ivanov2014}
\begin{eqnarray}\label{eq:20}
\hspace{-0.3in}\Big(\hat{k}_e - m_e(1 - \gamma) + i\frac{\alpha Z
  m_e}{k_e}\,\gamma^0\vec{\gamma}\cdot \vec{k}_e\Big) u_e(\vec{k}_e,
\sigma_e) = 0.
\end{eqnarray}
We normalize the wave function Eq.(\ref{eq:19}) in a standard way
$\bar{u}_e(\vec{k}_e,\sigma'_e) u_e(\vec{k}_e, \sigma_e) = 2
m_e\,\delta_{\sigma'_e \sigma_e}$.  Since $\gamma = O(\alpha^2)$,
keeping the contributions of order $O(\alpha)$ we have to set $\gamma
= 0$. The contribution of the Coulomb distortion to the
right--hand--side (r.h.s) of Eq.(\ref{eq:15}), multiplied by the
contribution of the phase--volume Eq.(\ref{eq:15}) is defined by the
trace
\begin{eqnarray}\label{eq:21}
\Phi_{\beta^-_c}(\vec{k}_e,\vec{k}_{\nu})\sum_{\rm pol}\frac{|M(n\to p
  e^-\nu_e)|^2}{32m^2_n G^2_F |V_{ud}|^2 E_e E_{\nu}} &&=: \frac{1 -
  \lambda^2}{1 + 3\lambda^2}\,i\,\frac{\alpha Z m_e}{k_e}\,\frac{{\rm
    tr}\{[(\vec{\sigma}\cdot \vec{k}_e),(\vec{\sigma}\cdot
    \vec{\xi}_e)](\vec{\sigma}\cdot \vec{k}_{\nu})\}}{4 E_e
  E_{\nu}} =\nonumber\\ &&= \frac{1 - \lambda^2}{1 +
  3\lambda^2}\,\frac{\alpha Z m_e}{k_e}\,\frac{\vec{\xi}_e\cdot
  (\vec{k}_e\times \vec{k}_{\nu})}{E_e E_{\nu}}.
\end{eqnarray}
We would like to emphasize that the contribution of the Coulomb
distortion of the Dirac wave function of a decay electron to the
correlation coefficient comes from the traces of $V\times V$ and
$A\times A$ products only, i.e. ${\rm tr}\{V\times V + A\times A\}
\sim (1 - \lambda^2)$. Thus, we get
\begin{eqnarray}\label{eq:22}
\hspace{-0.3in}&&\Phi_{\beta^-_c}(\vec{k}_e,\vec{k}_{\nu})\sum_{\rm
  pol}\frac{|M(n\to p e^-\nu_e)|^2}{32m^2_n G^2_F |V_{ud}|^2 E_e
  E_{\nu}} = (1 + 3\lambda^2)\tilde{\zeta}(E_e)\Big\{1 +
\tilde{a}(E_e)\frac{\vec{k}_e\cdot \vec{k}_{\nu}}{E_e E_{\nu}} +
\tilde{G}(E_e)\,\frac{\vec{\xi}_e\cdot \vec{k}_e}{E_e} +
\tilde{H}(E_e)\,\frac{\vec{\xi}_e \cdot
  \vec{k}_{\nu}}{E_{\nu}}\nonumber\\
\hspace{-0.3in}&&+ \tilde{K}_e(E_e)\,\frac{(\vec{\xi}_e\cdot
  \vec{k}_e)( \vec{k}_e\cdot \vec{k}_{\nu})}{(E_e + m_e)E_e E_{\nu}}+
\tilde{L}(E_e)\,\frac{\vec{\xi}_e\cdot (\vec{k}_e\times
  \vec{k}_{\nu})}{E_e E_{\nu}} - 3\,a_0\,
\frac{E_e}{M}\,\Big(\frac{(\vec{k}_e\cdot \vec{k}_{\nu})^2}{E^2_e
  E^2_{\nu}} - \frac{1}{3}\,\frac{k^2_e}{E^2_e}\,\Big) \nonumber\\
\hspace{-0.3in}&& + 3\,a_0\,
\frac{m_e}{M}\,\Big(\frac{(\vec{\xi}_e\cdot
  \vec{k}_{\nu})(\vec{k}_e\cdot \vec{k}_{\nu})}{E_e E^2_{\nu}} -
\frac{1}{3}\,\frac{\vec{\xi}_e\cdot \vec{k}_e}{E_e}\,\Big)+ 3\,a_0\,
\frac{1}{M}\,\Big(\frac{(\vec{\xi}_e\cdot \vec{k}_e)(\vec{k}_e\cdot
  \vec{k}_{\nu})^2}{(E_e + m_e)E_e E^2_{\nu}} - \frac{1}{3}\,(E_e -
m_e)\,\frac{\vec{\xi}_e\cdot \vec{k}_e}{E_e}\,\Big) \Big\}.
\end{eqnarray}
The correlation coefficient $\tilde{\zeta}(E_e)\tilde{L}(E_e)$
is equal to
\begin{eqnarray}\label{eq:23}
\hspace{-0.3in}\tilde{\zeta}(E_e)\tilde{L}(E_e) = \alpha\,\frac{
  m_e}{k_e}\,a_0,
\end{eqnarray}
where we have set $Z = 1$. Thus, the electron--energy and
electron--antineutrino angular distribution of the neutron
$\beta^-$--decay with polarized electron and unpolarized neutron and
proton is
 \begin{eqnarray}\label{eq:24}
\hspace{-0.3in}&&\frac{d^5
  \lambda_{\beta^-_c}(E_e,\vec{k}_e,\vec{\xi}_e, \vec{k}_{\nu})}{dE_e
  d\Omega_ed\Omega_{\nu}} = (1 +
3\lambda^2)\,\frac{G^2_F|V_{ud}|^2}{32\pi^5} (E_0 - E_e)^2 \sqrt{E^2_e
  - m^2_e}\, E_e\,F(E_e, Z = 1)\,\tilde{\zeta}(E_e) \Big\{1 +
\tilde{a}(E_e)\frac{\vec{k}_e\cdot \vec{k}_{\nu}}{E_e
  E_{\nu}}\nonumber\\
\hspace{-0.3in}&& + \tilde{G}(E_e)\,\frac{\vec{\xi}_e\cdot
  \vec{k}_e}{E_e} + \tilde{H}(E_e)\,\frac{\vec{\xi}_e \cdot
  \vec{k}_{\nu}}{E_{\nu}} + \tilde{K}_e(E_e)\,\frac{(\vec{\xi}_e\cdot
  \vec{k}_e)( \vec{k}_e\cdot \vec{k}_{\nu})}{(E_e + m_e)E_e E_{\nu}}+
\tilde{L}(E_e)\,\frac{\vec{\xi}_e\cdot (\vec{k}_e\times
  \vec{k}_{\nu})}{E_e E_{\nu}} - 3\,a_0\,
\frac{E_e}{M}\,\Big(\frac{(\vec{k}_e\cdot \vec{k}_{\nu})^2}{E^2_e
  E^2_{\nu}} - \frac{1}{3}\,\frac{k^2_e}{E^2_e}\,\Big) \nonumber\\
\hspace{-0.3in}&& + 3\,a_0\,
\frac{m_e}{M}\,\Big(\frac{(\vec{\xi}_e\cdot
  \vec{k}_{\nu})(\vec{k}_e\cdot \vec{k}_{\nu})}{E_e E^2_{\nu}} -
\frac{1}{3}\,\frac{\vec{\xi}_e\cdot \vec{k}_e}{E_e}\,\Big)+ 3\,a_0\,
\frac{1}{M}\,\Big(\frac{(\vec{\xi}_e\cdot \vec{k}_e)(\vec{k}_e\cdot
  \vec{k}_{\nu})^2}{(E_e + m_e)E_e E^2_{\nu}} - \frac{1}{3}\,(E_e -
m_e)\,\frac{\vec{\xi}_e\cdot \vec{k}_e}{E_e}\,\Big) \Big\}.
\end{eqnarray}
The radiative corrections to the correlation coefficients, defined
by the function $f_{\beta^-_c}(E_e, \mu)$, depend on the infrared
cut--off $\mu$. In order to remove such a dependence we have to add
the contribution of the neutron radiative $\beta^-$--decay
\cite{Sirlin1967}(see also \cite{Ivanov2013,Ivanov2017b}).

\section{Electron--Energy and electron--antineutrino angular 
distribution of neutron $\beta^-$--decay with polarized electron and
unpolarized neutron and proton to order $10^{-3}$}
\label{sec:spectrum}

\begin{figure}
\includegraphics[width=0.43\linewidth]{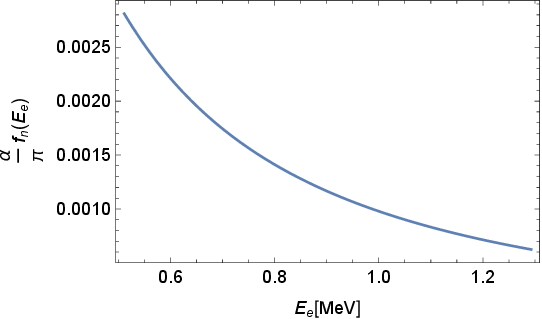} 
\includegraphics[width=0.43\linewidth]{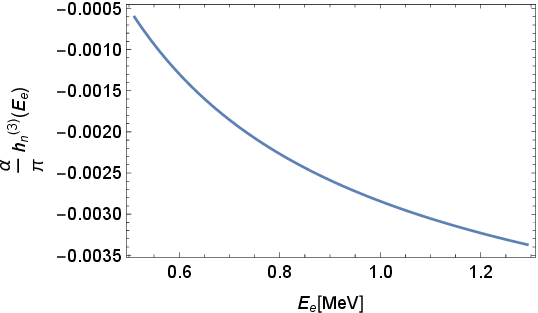} 
\includegraphics[width=0.43\linewidth]{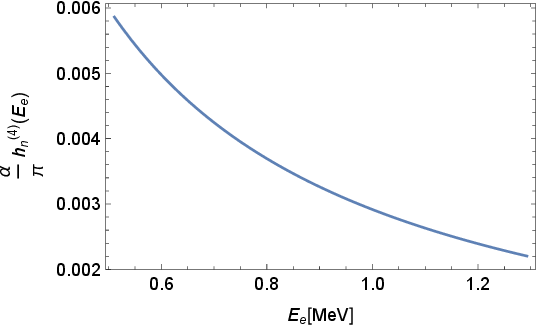} 
  \caption{Radiative corrections $(\alpha/\pi)\,f_n(E_e)$,
    $(\alpha/\pi)\,h^{(3)}_n(E_e)$ and $(\alpha/\pi)\,h^{(4)}_n(E_e)$
    to the correlation coefficients $G(E_e)$, $H(E_e)$ and $K_e(E_e)$
    of the electron--antineutrino energy and angular distribution
    Eq.(\ref{eq:25}).}
\label{fig:fig3}
\end{figure}

Summing the electron--energy and electron--antineutrino angular
distributions Eq.(\ref{eq:24}) and Eq.(\ref{eq:A.5}) in the Appendix
we obtain the electron--energy and electron--antineutrino angular
distribution of $\lambda_n = \lambda_{\beta^-_c}
+\lambda_{\beta^-\gamma}$ equal to
\begin{eqnarray*}
\hspace{-0.3in}&&\frac{d^5 \lambda_n(E_e,\vec{k}_e,\vec{\xi}_e,
  \vec{k}_{\nu})}{dE_e d\Omega_ed\Omega_{\nu}} = (1 +
3\lambda^2)\,\frac{G^2_F|V_{ud}|^2}{32\pi^5} (E_0 - E_e)^2 \sqrt{E^2_e
  - m^2_e}\, E_e\,F(E_e, Z = 1)\,\zeta(E_e)\,\Big\{1 +
a(E_e)\,\frac{\vec{k}_e\cdot \vec{k}_{\nu}}{E_e E_{\nu}}\nonumber\\
\end{eqnarray*}
\begin{eqnarray}\label{eq:25}
\hspace{-0.3in}&&+ G(E_e)\,\frac{\vec{\xi}_e\cdot \vec{k}_e}{E_e} +
H(E_e)\,\frac{\vec{\xi}_e \cdot \vec{k}_{\nu}}{E_{\nu}} +
K_e(E_e)\,\frac{(\vec{\xi}_e\cdot \vec{k}_e)( \vec{k}_e\cdot
  \vec{k}_{\nu})}{(E_e + m_e)E_e E_{\nu}}+
L(E_e)\,\frac{\vec{\xi}_e\cdot(\vec{k}_e \times
  \vec{k}_{\nu})}{E_eE_{\nu}} - 3\,a_0\,
\frac{E_e}{M}\,\Big(\frac{(\vec{k}_e\cdot \vec{k}_{\nu})^2}{E^2_e
  E^2_{\nu}} - \frac{1}{3}\,\frac{k^2_e}{E^2_e}\,\Big)\nonumber\\
\hspace{-0.3in}&& + 3\,a_0\,
\frac{m_e}{M}\,\Big(\frac{(\vec{\xi}_e\cdot
  \vec{k}_{\nu})(\vec{k}_e\cdot \vec{k}_{\nu})}{E_e E^2_{\nu}} -
\frac{1}{3}\,\frac{\vec{\xi}_e\cdot \vec{k}_e}{E_e}\,\Big) + 3\,a_0\,
\frac{1}{M}\,\Big(\frac{(\vec{\xi}_e\cdot \vec{k}_e)(\vec{k}_e\cdot
  \vec{k}_{\nu})^2}{(E_e + m_e)E_e E^2_{\nu}} - \frac{1}{3}\,(E_e -
m_e)\,\frac{\vec{\xi}_e\cdot \vec{k}_e}{E_e}\,\Big) \Big\}.
\end{eqnarray}
The correlation coefficients are equal to
\begin{eqnarray}\label{eq:26}
\hspace{-0.15in}\zeta(E_e) &=& \Big(1 +
\frac{\alpha}{\pi}\,g_n(E_e)\Big) + \frac{1}{M}\,\frac{1}{1 +
  3\lambda^2}\,\Big[- 2\,\lambda\Big(\lambda - (\kappa + 1)\Big)\,E_0
  + \Big(10 \lambda^2 - 4(\kappa + 1)\, \lambda + 2\Big)\,E_e\nonumber\\
\hspace{-0.15in}&& - 2 \lambda\,\Big(\lambda - (\kappa +
1)\Big)\,\frac{m^2_e}{E_e}\Big],\nonumber\\
\hspace{-0.15in}\zeta(E_e)\,a(E_e) &=& a_0 \Big(1 +
\frac{\alpha}{\pi}\,g_n(E_e) + \frac{\alpha}{\pi}\,f_n(E_e)\Big) +
\frac{1}{M}\,\frac{1}{1 + 3\lambda^2}\,\Big[2 \lambda \Big(\lambda -
  (\kappa + 1)\Big)E_0 - 4\lambda \Big(3\lambda - (\kappa + 1)\Big)
  E_e\Big],\nonumber\\
\hspace{-0.15in}\zeta(E_e)\,G(E_e) &=& - \Big(1 +
\frac{\alpha}{\pi}\,g_n(E_e) + \frac{\alpha}{\pi}\,f_n(E_e)\Big) +
\frac{1}{M}\,\frac{1}{1 + 3\lambda^2}\,\Big[2\lambda \Big(\lambda -
  (\kappa + 1)\Big)\,E_0 - \Big(10 \lambda^2 - 4(\kappa + 1)\,\lambda
  + 2\Big)E_e\Big]\nonumber\\
\hspace{-0.15in}\zeta(E_e) H(E_e) &=& \frac{m_e}{E_e}\,\Big\{- a_0
\Big(1 + \frac{\alpha}{\pi}\,g_n(E_e) +
\frac{\alpha}{\pi}\,h^{(3)}_n(E_e) \Big) + \frac{1}{M}\,\frac{1}{1 +
  3\lambda^2}\,\Big[ - 2\lambda\Big(\lambda - (\kappa + 1)\Big)
  E_0\nonumber\\
\hspace{-0.15in}&& + \Big(4\lambda^2 - 2 (\kappa + 1)\lambda - 2\Big)
E_e\Big]\Big\},\nonumber\\
\hspace{-0.15in}\zeta(E_e) K_e(E_e) &=& - a_0 \Big(1 +
\frac{\alpha}{\pi}\,g_n(E_e) + \frac{\alpha}{\pi}\,h^{(4)}_n(E_e)
\Big) + \frac{1}{M}\,\frac{1}{1 + 3\lambda^2}\,\Big[-
  2\lambda\Big(\lambda - (\kappa + 1)\Big) E_0 + 4\lambda\Big(3\lambda
  - (\kappa + 1)\Big)E_e\nonumber\\
\hspace{-0.15in}&& + \Big(8\lambda^2 - 2 (\kappa + 1)\lambda + 2\Big)
m_e\Big],\nonumber\\
\hspace{-0.15in}\zeta(E_e)L(E_e) &=& \alpha\,\frac{ m_e}{k_e}\,a_0
\end{eqnarray}
The radiative corrections of order $O(\alpha/\pi)$ to the correlation
coefficients are defined by the functions $g_n(E_e)$, $f_n(E_e)$ and
the functions $h^{(3)}_n(E_e)$ and $h^{(4)}_n(E_e)$. The functions
$g_n(E_e)$ and $f_n(E_e)$ have been calculated by Sirlin
\cite{Sirlin1967} and Shann \cite{Shann1971} (see also
\cite{Gudkov2006} and Appendices B, C, D, E and F in
Ref.\cite{Ivanov2013}), respectively. The contributions of the
electroweak--boson exchanges and QCD corrections to the function
$g_n(E_e)$ have been calculated in
\cite{Sirlin1986,Sirlin2004,Sirlin2006}. In turn, the radiative
corrections $(\alpha/\pi)\,h^{(3)}_n(E_e)$ and
$(\alpha/\pi)\,h^{(4)}_n(E_e)$ are calculated in Appendix A. The
functions $(\alpha/\pi)\,h^{(3)}_n(E_e)$ and
$(\alpha/\pi)\,h^{(4)}_n(E_e)$, together with the function
$(\alpha/\pi)f_n(E_e)$, are plotted in Fig.\,\ref{fig:fig3} in the
electron--energy region $m_e \le E_e \le E_0$.

\section{Correlation coefficients $a(E_e)$, $G(E_e)$, 
$H(E_e)$ and $K_e(E_e)$ to order $10^{-3}$}
\label{sec:order}

The correlation coefficients $a(E_e)$ and $G(E_e)$ have been
calculated in \cite{Ivanov2013} and \cite{Ivanov2017b},
respectively. They are equal to

\begin{eqnarray}\label{eq:27}
\hspace{-0.3in}a(E_e) &=& \Big(1 + \frac{\alpha}{\pi}\,f_n(E_e)\Big)
\Big\{a_0 + \frac{1}{M}\,\frac{1}{1 + 3\lambda^2}\,\Big[2 \lambda
  \Big(\lambda - (\kappa + 1)\Big)E_0 - 4\lambda \Big(3\lambda -
  (\kappa + 1)\Big) E_e\Big]\nonumber\\
\hspace{-0.3in}&+& \frac{1}{M}\, \frac{a_0}{1 + 3 \lambda^2}\,\Big[-
  \Big(10 \lambda^2 - 4(\kappa + 1)\,\lambda + 2\Big)\,E_e + \Big(2
  \lambda^2 - 2(\kappa + 1)\,\lambda\Big)\,\Big(E_0 +
  \frac{m^2_e}{E_e}\Big)\Big]\Big\},\nonumber\\
\hspace{-0.3in}G(E_e) &=&  - \Big(1 +
\frac{\alpha}{\pi}\,f_n(E_e)\Big)\,\Big(1 + \frac{1}{M}\,\frac{1}{1 +
  3\lambda^2}\,\Big(2 \lambda^2 - 2 (\kappa +
1)\,\lambda\Big)\,\frac{m^2_e}{E_e}\Big).
\end{eqnarray}
For the correlation coefficients $H(E_e)$ and $K_e(E_e)$ we obtain the
following expressions
\begin{eqnarray}\label{eq:28}
\hspace{-0.3in}H(E_e) &=& \Big(1 +
\frac{\alpha}{\pi}\,h^{(3)}_n(E_e)\Big)\,\frac{m_e}{E_e}\,\Big\{ - a_0
+ \frac{1}{M}\,\frac{1}{1 + 3\lambda^2}\,\Big[ - 2\lambda\Big(\lambda
  - (\kappa + 1)\Big) E_0 + \Big(4\lambda^2 - 2 (\kappa + 1)\lambda -
  2\Big) E_e\Big]\nonumber\\
\hspace{-0.3in}&-& \frac{1}{M}\, \frac{a_0}{1 + 3 \lambda^2}\,\Big[-
  \Big(10 \lambda^2 - 4(\kappa + 1)\,\lambda + 2\Big)\,E_e + \Big(2
  \lambda^2 - 2(\kappa + 1)\,\lambda\Big)\,\Big(E_0 +
  \frac{m^2_e}{E_e}\Big)\Big]\Big\}
\end{eqnarray}
and 
\begin{eqnarray}\label{eq:29}
\hspace{-0.3in}K_e(E_e) &=& \Big(1 +
\frac{\alpha}{\pi}\,h^{(4)}_n(E_e)\Big)\,\Big\{ - a_0 +
\frac{1}{M}\,\frac{1}{1 + 3\lambda^2}\,\Big[- 2\lambda\Big(\lambda -
  (\kappa + 1)\Big) E_0 + 4\lambda\Big(3\lambda - (\kappa +
  1)\Big)E_e\nonumber\\
\hspace{-0.3in}&+& \Big(8\lambda^2 - 2 (\kappa + 1)\lambda + 2\Big)
m_e\Big] - \frac{1}{M}\, \frac{a_0}{1 + 3 \lambda^2}\,\Big[- \Big(10
  \lambda^2 - 4(\kappa + 1)\,\lambda + 2\Big)\,E_e \nonumber\\
\hspace{-0.3in}&+& \Big(2 \lambda^2 - 2(\kappa +
1)\,\lambda\Big)\,\Big(E_0 + \frac{m^2_e}{E_e}\Big)\Big]\Big\}
\end{eqnarray}
The obtained correlation coefficients are calculated to order
$10^{-3}$, taking into account the complete set of corrections of
order $O(E_e/M)$ and $O(\alpha/\pi)$, caused by the weak magnetism,
proton recoil and one--photon exchanges, respectively.

\section{Wilkinson's corrections}
\label{sec:wilkinson}

According to Wilkinson \cite{Wilkinson1982}, the higher order
corrections with respect to those calculated in section
\ref{sec:order} should be caused by i) the proton recoil in the
Coulomb electron--proton final--state interaction, ii) the finite
proton radius, iii) the proton--lepton convolution and iv) the
higher--order {\it outer} radiative corrections.

\begin{figure}
\includegraphics[width=0.45\linewidth]{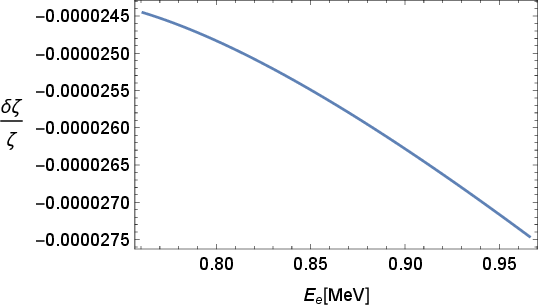} 
\includegraphics[width=0.43\linewidth]{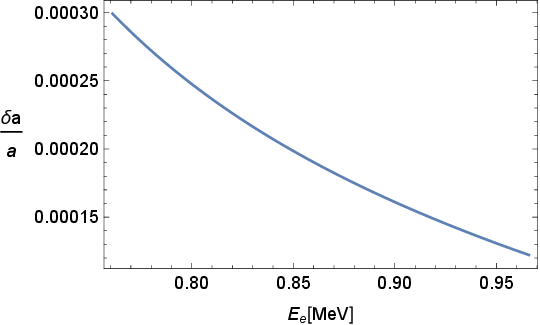}
\includegraphics[width=0.43\linewidth]{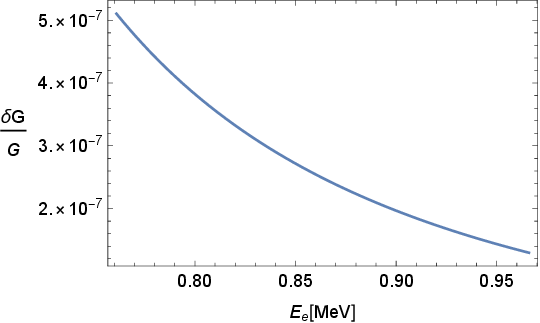}
\includegraphics[width=0.45\linewidth]{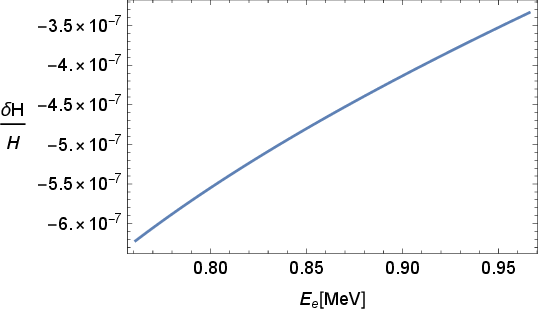}
\includegraphics[width=0.43\linewidth]{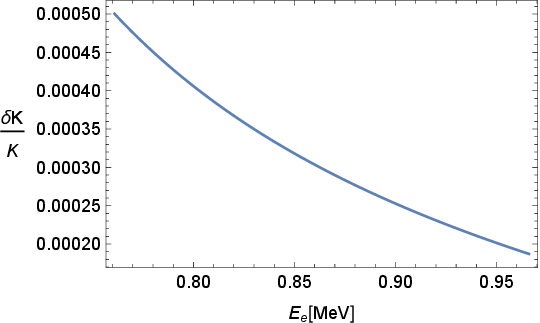}
  \caption{Relative corrections to the correlation coefficients
    $\zeta(E_e)$,  $a(E_e)$, $G(E_e)$, $H(E_e)$ and $K_e(E_e)$ induced
    by the proton recoil to the Fermi function, caused by the Coulomb
    electron--proton final--state interaction and calculated for the
    experimentally observable electron energy region $0.761\,{\rm MeV}
    \le E_e \le 0.966\,{\rm MeV}$ \cite{Ivanov2013}.}
\label{fig:fig4}
\end{figure}

The relative corrections to the correlation coefficients $\zeta(E_e)$,
$a(E_e)$, $G(E_e)$, $H(E_e)$ and $K_e(E_e)$, caused by the proton recoil
in the final state electron--proton Coulomb interactions, are equal to
\begin{eqnarray}\label{eq:30}
\hspace{-0.3in}\frac{\delta \zeta(E_e)}{\zeta(E_e)} &=&- \frac{\pi
  \alpha}{\beta}\,\frac{E_e}{M} - \frac{1}{3}\,\frac{1 - \lambda^2}{1
  + 3\lambda^2}\,\frac{\pi \alpha}{\beta}\,\frac{E_0 -
  E_e}{M},\nonumber\\
\hspace{-0.3in}\frac{\delta a(E_e)}{a(E_e)} &=&\frac{1}{3}\,\frac{1 -
  \lambda^2}{1 + 3\lambda^2}\,\frac{\pi \alpha}{\beta}\,\frac{E_0 -
  E_e}{M} - \frac{1 + 3\lambda^2}{1 - \lambda^2}\,
\frac{\pi \alpha}{\beta^3}\,\frac{E_0 - E_e}{M},\nonumber\\
\hspace{-0.3in}\frac{\delta G(E_e)}{G(E_e)} &=& - \frac{1}{3}\,\frac{1
  - \lambda^2}{1 + 3\lambda^2}\, (1 - \beta^2)\,\frac{\pi
  \alpha}{\beta^3}\,\frac{E_0 - E_e}{M},\nonumber\\
\hspace{-0.3in}\frac{\delta H(E_e)}{H(E_e)} &=& \frac{1}{3}\,\frac{1
  - \lambda^2}{1 + 3\lambda^2}\,\frac{\pi
  \alpha}{\beta}\,\frac{E_0 - E_e}{M},\nonumber\\
\hspace{-0.3in}\frac{\delta K_e(E_e)}{K_e(E_e)} &=&
\frac{1}{3}\,\frac{1 - \lambda^2}{1 + 3\lambda^2}\,\frac{\pi
  \alpha}{\beta}\,\frac{E_0 - E_e}{M} - \frac{1 + 3\lambda^2}{1 -
  \lambda^2}\,\frac{\pi \alpha}{\beta^3}\,\frac{E_0 - E_e}{M}\,\big(1
+ \sqrt{1 - \beta^2}\,\big).
\end{eqnarray}
\begin{table}[h]
\begin{tabular}{|c|c|c|}
\hline $E_e = 0.761\,{\rm MeV}$ & $\delta X(E_e)/X(E_e)$ & $E_e =
0.966\,{\rm MeV}$ \\ \hline $- 2.5\times 10^{-5}$ & $\ge \delta
\zeta(E_e)/\zeta(E_e) \ge $ & $- 2.8 \times 10^{-5}$ \\ \hline $+ 3.0
\times 10^{-4}$ & $ \ge \delta a(E_e)/a(E_e) \ge $ & $+ 1.1 \times
10^{-4}$\\ \hline $+ 5.1\times 10^{-7}$ & $ \ge \delta G(E_e)/G(E_e)
\ge$ & $+ 1.3 \times 10^{-7}$ \\ \hline $- 6.2\times 10^{-7} $ & $ \le
\delta H(E_e)/H(E_e) \le$ & $ - 3.3 \times 10^{-7}$\\ \hline $+
5.0\times 10^{-4} $ & $ \ge \delta K_e(E_e)/K_e(E_e) \ge $ & $+ 1.9
\times 10^{-4}$\\ \hline
\end{tabular} 
\caption{Wilkinson's corrections, induced by the change of the Fermi
  function caused by the electron--proton final--state Coulomb
  interaction, in the energy region $0.761\,{\rm MeV} \le E_e \le
  0.966\,{\rm MeV}$.}
\end{table}
In the experimental electron energy region $0.761\,{\rm MeV} \le E_e
\le 0.966\,{\rm MeV}$ the corrections Eq.(\ref{eq:30}) are plotted in
Fig\,\ref{fig:fig4} and take the values adduced in Table I. The proton
recoil corrections to the correlation coefficient $a(E_e)$, caused by
the electron--proton final--state Coulomb interactions, are of order
$10^{-4}$ and should be taken into account for the analysis of the
experimental data on searches of contributions of interactions beyond
the SM at the level of $10^{-4}$ \cite{Abele2016}.

In turn, Wilkinson's corrections, caused by ii) the finite proton
radius, iii) the proton--lepton convolution and iv) the higher--order
{\it outer} radiative corrections, retain their expression for
calculated in \cite{Ivanov2017b} and the order $|\delta
\zeta(E_e)/\zeta(E_e)| \sim 10^{-5}$, $|\delta a(E_e)/a(E_e)| \sim
|\delta K_e(E_e)/K_e(E_e)| \sim 10^{-4}$, and $|\delta G(E_e)/G(E_e)|
\sim |\delta H(E_e)/H(E_e)| \sim 10^{-7}$, respectively.

\section{Electron--energy and electron--antineutrino angular 
distribution beyond the SM}
\label{sec:bsm}

For the calculation of contributions of interactions beyond the SM we
use the effective low--energy Hamiltonian of weak nucleon--lepton
four--fermion local interactions, taking into account all
phenomenological couplings beyond the SM
\cite{Lee1956}--\cite{Gardner2013} in the notations of
\cite{Ivanov2013,Ivanov2017d}:
\begin{eqnarray}\label{eq:31}
{\cal H}_W(x) &=&
\frac{G_F}{\sqrt{2}}\,V_{ud}\Big\{[\bar{\psi}_p(x)\gamma_{\mu}\psi_n(x)]
     [\bar{\psi}_e(x)\gamma^{\mu}(C_V + \bar{C}_V
       \gamma^5)\psi_{\nu_e}(x)] +
     [\bar{\psi}_p(x)\gamma_{\mu}\gamma^5\psi_n(x)][\bar{\psi}_e(x)
       \gamma^{\mu}(\bar{C}_A + C_A \gamma^5)\psi_{\nu_e}(x)]
     \nonumber\\ &+& [\bar{\psi}_p(x)\psi_n(x)][\bar{\psi}_e(x)(C_S +
       \bar{C}_S \gamma^5)\psi_{\nu_e}(x)] + [\bar{\psi}_p(x) \gamma
       ^5 \psi_n(x)][\bar{\psi}_e(x)(C_P + \bar{C}_P
       \gamma^5)\psi_{\nu_e}(x)]\nonumber\\ &+&\frac{1}{2}
                 [\bar{\psi}_p(x)\sigma^{\mu\nu}\psi_n(x)]
                 [\bar{\psi}_e(x)\sigma_{\mu\nu} (C_T + \bar{C}_T
                   \gamma^5)\psi_{\nu_e}(x) \Big\}.
\end{eqnarray}
This is the most general form of the effective low--energy weak
interactions, where the phenomenological coupling constants $C_i$ and
$\bar{C}_i$ for $i = V, A, S, P$ and $T$ can be induced by the
left--handed and right--handed hadronic and leptonic currents
\cite{Lee1956}--\cite{Severijns2006}. They are related to the
phenomenological coupling constants, analogous to those which were
introduced by Herczeg \cite{Herczeg2001}, as follows
\begin{eqnarray}\label{eq:32}
\hspace{-0.3in} C_V &=&1 + a^h_{LL} + a^h_{LR} + a^h_{RR} +
a^h_{RL}\quad,\quad \bar{C}_V = - 1 - a^h_{LL} -
a^h_{LR} + a^h_{RR} + a^h_{RL},\nonumber\\
\hspace{-0.3in} C_A &=& -\lambda + a^h_{LL} - a^h_{LR} + a^h_{RR} -
a^h_{RL}\quad,\quad \bar{C}_A = \lambda - a^h_{LL} + a^h_{LR} + a^h_{RR} -
a^h_{RL},\nonumber\\
\hspace{-0.3in}C_S &=& A^h_{LL} + A^h_{LR} +
A^h_{RR} + A^h_{RL} \quad,\quad \bar{C}_S =  - A^h_{LL} -
A^h_{LR} + A^h_{RR} + A^h_{RL},\nonumber\\
\hspace{-0.3in}C_P &=& - A^h_{LL} + A^h_{LR} + A^h_{RR} -
A^h_{RL}\quad,\quad \bar{C}_P = A^h_{LL} - A^h_{LR} + A^h_{RR} -
A^h_{RL},\nonumber\\
\hspace{-0.3in} C_T &=& 2( \alpha^h_{LL} + \alpha^h_{RR})\quad,\quad
\bar{C}_T =  2( - \alpha^h_{LL} + \alpha^h_{RR}),
\end{eqnarray}
where the index $h$ means that the phenomenological coupling
constants are introduced at the {\it hadronic} level but not at the
{\it quark} level as it has been done by Herczeg
\cite{Herczeg2001}. In the SM the phenomenological coupling constants
$C_i$ and $\bar{C}_i$ for $i = V, A, S, P$ and $T$ are equal to $C_S =
\bar{C}_S = C_P = \bar{C}_P = C_T = \bar{C}_T = 0$, $C_V = -
\,\bar{C}_V = 1$ and $C_A = - \,\bar{C}_A = - \lambda$
\cite{Ivanov2013}.  The phenomenological coupling constants
$a^h_{ij}$, $A^h_{ij}$ and $\alpha^h_{jj}$ for $i(j) = L$ or $R$ are
induced by interactions beyond the SM. 

The contribution of interactions beyond the SM, given by the
Hamiltonian of weak interactions Eq.(\ref{eq:6}), to the amplitude of
the neutron $\beta^-$--decay, calculated to leading order in the large
nucleon mass expansion, takes the form
\begin{eqnarray}\label{eq:33}
\hspace{-0.3in} M(n \to p e^- \bar{\nu}_e) &=& -\,2m_n\,
\frac{G_F}{\sqrt{2}}\,V_{ud}\,\Big\{[\varphi^{\dagger}_p\varphi_n]
     [\bar{u}_e \gamma^0(C_V + \bar{C}_V \gamma^5) v_{\bar{\nu}}] -
     [\varphi^{\dagger}_p\vec{\sigma}\,\varphi_n]\cdot [\bar{u}_e
       \vec{\gamma}\,(\bar{C}_A + C_A \gamma^5)
       v_{\bar{\nu}}]\nonumber\\
\hspace{-0.3in}&&+ [\varphi^{\dagger}_p\varphi_n][\bar{u}_e (C_S +
  \bar{C}_S \gamma^5) v_{\bar{\nu}}] +
       [\varphi^{\dagger}_p\vec{\sigma}\,\varphi_n]\cdot [\bar{u}_e
         \gamma^0 \vec{\gamma}\,(\bar{C}_T + C_T \gamma^5)
         v_{\bar{\nu}}]\Big\}.
\end{eqnarray}
The hermitian conjugate amplitude is
\begin{eqnarray}\label{eq:34}
\hspace{-0.3in} M^{\dagger}(n \to p e^- \bar{\nu}_e) &=& -\,2m_n\,
\frac{G_F}{\sqrt{2}}\,V^*_{ud}\Big\{[\varphi^{\dagger}_n\varphi_p]
     [\bar{v}_{\bar{\nu}} \gamma^0(C^*_V + \bar{C}^*_V \gamma^5) u_e]
     - [\varphi^{\dagger}_n\vec{\sigma}\,\varphi_p]\cdot
     [\bar{v}_{\bar{\nu}} \vec{\gamma}\,(\bar{C}^*_A + C^*_A \gamma^5)
       u_e] \nonumber\\
\hspace{-0.3in}&&+ [\varphi^{\dagger}_n\varphi_p][\bar{v}_{\bar{\nu}}
  (C^*_S - \bar{C}^*_S \gamma^5) u_e] -
       [\varphi^{\dagger}_n\vec{\sigma}\,\varphi_p]\cdot
       [\bar{v}_{\bar{\nu}} \gamma^0 \vec{\gamma}\,(\bar{C}^*_T -
         C^*_T \gamma^5) u_e]\Big\}.
\end{eqnarray}
The contributions of interactions with the strength, defined by the
phenomenological coupling constants $C_P$ and $\bar{C}_P$, may appear
only of order $O(C_P E_e/M)$ and $O(\bar{C}_P E_e/M)$ and can be
neglected to leading order in the large nucleon mass expansion. We
have also neglected the contributions of the neutron--proton mass
difference. The squared absolute value of the amplitude
Eq.(\ref{eq:8}), summed over polarizations of massive fermions, is
equal to
\begin{eqnarray}\label{eq:35}
\hspace{-0.15in}&& \sum_{\rm pol.}\frac{|M(n \to p e^-
  \bar{\nu}_e)|^2}{8m^2_n G^2_F|V_{ud}|^2 E_{\nu}E_e} =
\Big\{\frac{1}{2}\Big(|C_V|^2 + |\bar{C}_V|^2 + 3|C_A|^2 +
3|\bar{C}_A|^2 + |C_S|^2 + |\bar{C}_S|^2 + 3 |C_T|^2 +
3|\bar{C}_T|^2\Big)\nonumber\\
\hspace{-0.15in}&& + \frac{m_e}{E_e}\,{\rm Re}\Big(C_VC^*_S +
\bar{C}_V\bar{C}^*_S - 3 C_AC^*_T - 3 \bar{C}_A\bar{C}^*_T\Big) +
\frac{\vec{k}_e\cdot
  \vec{k}_{\nu}}{E_eE_{\nu}}\,\frac{1}{2}\Big(|C_V|^2 + |\bar{C}_V|^2
- |C_A|^2 - |\bar{C}_A|^2 - |C_S|^2\nonumber\\
\hspace{-0.15in}&& - |\bar{C}_S|^2 + |C_T|^2 + |\bar{C}_T|^2 \Big) +
\frac{\vec{\xi}_e\cdot \vec{k}_e}{E_e}\,{\rm Re}\Big(C_V\bar{C}^*_V +
3 C_A\bar{C}^*_A - C_S\bar{C}^*_S - 3 C_T\bar{C}^*_T \Big) +
\frac{\vec{\xi}_e\cdot \vec{k}_{\nu}}{E_{\nu}}\,{\rm Re}\Big( C_V
\bar{C}^*_S + \bar{C}_V C^*_S \nonumber\\
\hspace{-0.15in}&&+ C_A \bar{C}^*_T + \bar{C}_A C^*_T +
\frac{m_e}{E_e}(C_V \bar{C}^*_V - C_A \bar{C}^*_A + C_S \bar{C}^*_S -
C_T \bar{C}^*_T)\Big) + \frac{(\vec{\xi}_e\cdot
  \vec{k}_e)(\vec{k}_e\cdot \vec{k}_{\nu})}{(E_e + m_e)E_e
  E_{\nu}}\,{\rm Re}\Big(C_V \bar{C}^*_V - C_A \bar{C}^*_A\nonumber\\
\hspace{-0.15in}&& + C_S \bar{C}^*_S - C_T \bar{C}^*_T - C_V
\bar{C}^*_S - \bar{C}_V C^*_S - C_A \bar{C}^*_T - \bar{C}_A C^*_T\Big)
+ \frac{\vec{\xi}_e\cdot (\vec{k}_e \times
  \vec{k}_{\nu})}{E_eE_{\nu}}\,{\rm Im}\Big(C_S C^*_V + \bar{C}_S
\bar{C}^*_V +  C_T C^*_A\nonumber\\
\hspace{-0.15in}&& + \bar{C}_T \bar{C}^*_A \Big)\Big\}.
\end{eqnarray}
The structure of the correlation coefficients in Eq.(\ref{eq:35})
agrees well with the structure of the corresponding expressions
obtained in \cite{Jackson1957}. In the linear approximation for
coupling constants of vector and axial--vector interactions beyond the
SM \cite{Ivanov2013} we get
\begin{eqnarray}\label{eq:36}
\hspace{-0.3in}&& \sum_{\rm pol.}\frac{|M(n \to p e^-
  \bar{\nu}_e)|^2}{8m^2_n G^2_F|V_{ud}|^2 E_{\nu}E_e\,(1 + 3
  \lambda^2)} = \,\Big\{\Big[1 + \frac{1}{2}\,\frac{1}{1 +
      3\lambda^2}\,( |C_S|^2 + |\bar{C}_S|^2 + 3 |C_T|^2 +
    3|\bar{C}_T|^2)\Big]\nonumber\\
\hspace{-0.3in}&& + \frac{m_e}{E_e}\,\frac{1}{1 + 3\lambda^2}\,{\rm
  Re}\Big((C_S - \bar{C}_S) + 3 \lambda\,(C_T -\bar{C}_T)\Big) +
\frac{\vec{k}_e\cdot \vec{k}_{\nu}}{E_eE_{\nu}}\,\Big[a_0 -
\frac{1}{2}\,\frac{1}{1 + 3\lambda^2}(|C_S|^2 + |\bar{C}_S|^2 -
|C_T|^2 - |\bar{C}_T|^2)\Big] \nonumber\\
\hspace{-0.15in}&& + \frac{\vec{k}_e\cdot \vec{\xi}_e}{E_e}\,\Big[- 1
  - \frac{1}{1 + 3\lambda^2}\,{\rm Re}\Big(C_S\bar{C}^*_S + 3
  C_T\bar{C}^*_T\Big) \Big] + \frac{\vec{\xi}_e\cdot
  \vec{k}_{\nu}}{E_{\nu}}\,\Big[- \frac{m_e}{E_e}\,a_0 - \frac{1}{1 +
    3\lambda^2}{\rm Re}\Big((C_S - \bar{C}_S) - \lambda (C_T -
  \bar{C}_T)\Big)\nonumber\\
\hspace{-0.15in}&& + \frac{m_e}{E_e}\,\frac{1}{1 + 3\lambda^2}{\rm
  Re}\Big(C_S \bar{C}^*_S - C_T \bar{C}^*_T\Big)\Big] +
\frac{(\vec{\xi}_e\cdot \vec{k}_e)(\vec{k}_e\cdot \vec{k}_{\nu})}{(E_e
  + m_e)E_e E_{\nu}}\,\Big[- a_0 + \frac{1}{1 +
  3\lambda^2}{\rm Re}\Big((C_S - \bar{C}_S) - \lambda (C_T -
\bar{C}_T)\Big)\nonumber\\
\hspace{-0.15in}&& + \frac{1}{1 + 3\lambda^2}{\rm Re}\Big(C_S
\bar{C}^*_S - C_T \bar{C}^*_T\Big)\Big] + \frac{\vec{\xi}_e\cdot
  (\vec{k}_e \times \vec{k}_{\nu})}{E_eE_{\nu}}\,\frac{1}{1 +
  3\lambda^2}{\rm Im}\Big((C_S - \bar{C}_S) - \lambda (C_T -
\bar{C}_T)\Big)\Big\},
\end{eqnarray}
where we have replaced $C_j$ and $\bar{C}_j$ with $j = V,A$ by $C_V =
1 + \delta C_V$, $\bar{C}_V = - 1 + \delta \bar{C}_V$, $C_A = -
\lambda + \delta C_A$ and $\bar{C}_A = \lambda + \delta \bar{C}_A$
\cite{Ivanov2013} and neglected also the contributions of the products
$\delta C_j C_k$, $\delta \bar{C}_j C_k$ and so on for $j = V,A$ and
$k = S,T$. Following \cite{Bhattacharya2012,Cirigliano2013}(see also
\cite{Ivanov2013}) we have absorbed the contributions the vector and
  axial vector interactions beyond the SM by the axial coupling
  constant $\lambda$ and the CKM matrix element $V_{ud}$.

Thus, the electron--energy and electron--antineutrino angular
distribution Eq.(\ref{eq:1}), taking into account the contributions of
interactions beyond the SM, can be transcribed into the form
\begin{eqnarray}\label{eq:37}
&&\frac{d^3 \lambda_n(E_e,\vec{k}_e, \vec{\xi}_n,\vec{\xi}_e)}{dE_e
    d\Omega_e} = (1 + 3 \lambda^2)\,\frac{G^2_F|V_{ud}|^2}{8\pi^4}
  (E_0 - E_e)^2 \sqrt{E^2_e - m^2_e}\, E_e F(E_e, Z = 1) \zeta^{(\rm
    SM)}(E_e) \nonumber\\ &&\times \Big(1 + \zeta^{(\rm
    BSM)}(E_e)\Big)\Big\{1 + b\,\frac{m_e}{E_e}+ a_{\rm
    eff}(E_e)\,\frac{\vec{k}_e\cdot \vec{k}_{\nu}}{E_e E_{\nu}} +
  G_{\rm eff}(E_e)\,\frac{\vec{\xi}_e\cdot \vec{k}_e}{E_e} + H_{\rm
    eff}(E_e)\,\frac{\vec{\xi}_e \cdot
    \vec{k}_{\nu}}{E_{\nu}}\nonumber\\
\hspace{-0.15in}&&+ K_{e,\rm eff}(E_e)\,\frac{(\vec{\xi}_e\cdot
  \vec{k}_e)( \vec{k}_e\cdot \vec{k}_{\nu})}{(E_e + m_e)E_e E_{\nu}}+
L_{\rm eff}(E_e)\,\frac{\vec{\xi}_e\cdot(\vec{k}_e \times
  \vec{k}_{\nu})}{E_eE_{\nu}} - 3\,a_0\, \frac{E_e}{M}\,
\Big(\frac{(\vec{k}_e\cdot \vec{k}_{\nu})^2}{E^2_e E^2_{\nu}} -
\frac{1}{3}\,\frac{k^2_e}{E^2_e}\,\Big) \nonumber\\
\hspace{-0.15in}&&+ 3\,a_0\,
\frac{m_e}{M}\,\Big(\frac{(\vec{\xi}_e\cdot
  \vec{k}_{\nu})(\vec{k}_e\cdot \vec{k}_{\nu})}{E_e E^2_{\nu}} -
\frac{1}{3}\,\frac{\vec{\xi}_e\cdot \vec{k}_e}{E_e}\,\Big) + 3\,a_0\,
\frac{1}{M}\,\Big(\frac{(\vec{\xi}_e\cdot \vec{k}_e)(\vec{k}_e\cdot
  \vec{k}_{\nu})^2}{(E_e + m_e)E_e E^2_{\nu}} - \frac{1}{3}\,(E_e -
m_e)\,\frac{\vec{\xi}_e\cdot \vec{k}_e}{E_e}\,\Big) \Big\}.
\end{eqnarray}
where the indices ``SM'' and ``BSM'' mean ``Standard Model'' and
``Beyond Standard Model'', respectively.  The correlation coefficient
$\zeta^{(\rm SM)}(E_e)$ is given in Eq.(\ref{eq:25}). The Fierz
interference term $b$ and the correlation coefficients $X_{\rm
  eff}(E_e)$ with $X = a, G, H$ and $K_e$ are defined by
\begin{eqnarray}\label{eq:38}
b &=& \frac{b_F}{\displaystyle 1 + \zeta^{(\rm BSM)}(E_e)}\quad,\quad
a_{\rm eff}(E_e) = \frac{a^{(\rm SM)}(E_e) + a^{(\rm
    BSM)}(E_e)}{\displaystyle 1 + \zeta^{(\rm
    BSM)}(E_e)},\nonumber\\ G_{\rm eff}(E_e) &=& \frac{G^{(\rm
    SM)}(E_e) + G^{(\rm BSM)}(E_e)}{\displaystyle 1 + \zeta^{(\rm
    BSM)}(E_e)}\quad,\quad H_{e,\rm eff}(E_e) = \frac{H^{(\rm
    SM)}_e(E_e) + H^{(\rm BSM)}_e(E_e)}{\displaystyle 1 + \zeta^{(\rm
    BSM)}(E_e)},\nonumber\\ K_{\rm eff}(E_e) &=& \frac{K^{(\rm
    SM)}(E_e) + K^{(\rm BSM)}(E_e)}{\displaystyle 1 + \zeta^{(\rm
    BSM)}(E_e)}\quad,\quad L_{\rm eff}(E_e) = \frac{L^{(\rm
    SM)}(E_e) + L^{(\rm BSM)}(E_e)}{\displaystyle 1 + \zeta^{(\rm
    BSM)}(E_e)},
\end{eqnarray}
where the correlation coefficients with index ``SM'' are adduced in
Eqs.(\ref{eq:27}) - (\ref{eq:29}). They should be also supplemented by
Wilkinson's corrections Eq.(\ref{eq:30}) and those obtained in
\cite{Ivanov2017b} (see Chapter III of Ref.\cite{Ivanov2017b}).  The
correlation coefficients $b_F$ and the correlation coefficients with
index ``BSM'' are given by
\begin{eqnarray}\label{eq:39}
b_F &=& \frac{1}{1 + 3\lambda^2}\,{\rm Re}\Big((C_S - \bar{C}_S) + 3
\lambda\,(C_T -\bar{C}_T)\Big),\nonumber\\ 
\hspace{-0.3in}\zeta^{(\rm BSM)}(E_e) &=& \frac{1}{2}\,\frac{1}{1 +
  3\lambda^2}\,\Big( |C_S|^2 + |\bar{C}_S|^2 + 3 |C_T|^2 +
3|\bar{C}_T|^2\Big),\nonumber\\ 
a^{(\rm BSM)}(E_e) &=& -
\frac{1}{2}\,\frac{1}{1 + 3\lambda^2}\Big(|C_S|^2 + |\bar{C}_S|^2 -
|C_T|^2 - |\bar{C}_T|^2\Big),\nonumber\\ G^{(\rm BSM)}(E_e) &=& -
\frac{1}{1 + 3\lambda^2}\,{\rm Re}\Big(C_S\bar{C}^*_S + 3
C_T\bar{C}^*_T\Big),\nonumber\\ 
H^{(\rm BSM)}(E_e) &=&
\frac{m_e}{E_e}\,\frac{1}{1 + 3\lambda^2}{\rm Re}\Big(C_S \bar{C}^*_S
- C_T \bar{C}^*_T\Big) - \frac{1}{1 + 3\lambda^2}{\rm Re}\Big((C_S -
\bar{C}_S) - \lambda (C_T - \bar{C}_T)\Big) \nonumber\\ 
K^{(\rm BSM)}_e(E_e)&=& \frac{1}{1 + 3\lambda^2}{\rm Re}\Big(C_S
\bar{C}^*_S - C_T \bar{C}^*_T\Big) + \frac{1}{1 + 3\lambda^2}{\rm
  Re}\Big((C_S - \bar{C}_S) - \lambda (C_T -
\bar{C}_T)\Big),\nonumber\\ 
L^{(\rm BSM)}_e(E_e)&=& \frac{1}{1 +
  3\lambda^2}{\rm Im}\Big((C_S - \bar{C}_S) - \lambda (C_T -
\bar{C}_T)\Big).
\end{eqnarray}
The correlation coefficient $X_{\rm eff}(E_e)$ with $X = a, G, H$ and
$K_e$ are given in the form suitable for the analysis of experimental
data of experiments on the searches of interactions beyond the SM
\cite{Abele2016}. The structure of the correlation coefficients in
Eq.(\ref{eq:39}) agrees well with the structure of corresponding
expressions calculated in \cite{Jackson1957}.  The averaged values of
the correlation coefficients $X_{\rm eff}(E_e)$ with $X = a, G, H$ and
$K_e$ can be obtained with the electron--energy density
\cite{Ivanov2017d}
\begin{eqnarray}\label{eq:40}
\rho_e(E_e) = \rho^{(\rm SM)}_e(E_e)\,\big(1 + \zeta^{(\rm
  BSM)}(E_e)\big) = \rho^{(\rm SM)}_e(E_e)\,\Big( 1 +
\frac{1}{2}\,\frac{1}{1 + 3\lambda^2}\,( |C_S|^2 + |\bar{C}_S|^2 + 3
|C_T|^2 + 3|\bar{C}_T|^2)\Big),
\end{eqnarray}
where the electron--energy density $\rho^{(\rm SM)}_e(E_e)$ is defined
by Eq.(D-59) of Ref.\cite{Ivanov2013}.

\section{G--odd correlations}
\label{sec:gparity}

The $G$--parity transformation, i.e. $G = C\,e^{\,i \pi I_2}$, where
$C$ and $I_2$ are the charge conjugation and isospin operators, was
introduced by Lee and Yang \cite{Lee1956a} as a symmetry of strong
interactions. According to the $G$--transformation properties of
hadronic currents, Weinberg divided hadronic currents into two
classes, which are $G$--even first class and $G$--odd second class
currents \cite{Weinberg1958}, respectively. Following Weinberg
\cite{Weinberg1958}, Gardner and Zhang \cite{Gardner2001}, and Gardner
and Plaster \cite{Gardner2013} the $G$--odd contribution
to the matrix element of the hadronic $n \to p$ transition in the $V -
A$ theory of weak interactions can be taken in the following form
\begin{eqnarray}\label{eq:41}
\langle
p(\vec{k}_p,\sigma_p)|J^{(+)}_{\mu}(0)|n(\vec{k}_n,\sigma_n)\rangle_{G
  - \rm odd} =
\bar{u}_p(\vec{k}_p,\sigma_p)\Big(\frac{q_{\mu}}{M}\,f_3(0) +
i\frac{1}{M}\,\sigma_{\mu\nu}\gamma^5 q^{\nu}g_2(0)\Big)\,u_n(\vec{k}_n,
\sigma_n),
\end{eqnarray}
where $J^{(+)}_{\mu}(0) = V^{(+)}_{\mu}(0) - A^{(+)}_{\mu}(0)$,
$\bar{u}_p(\vec{k}_p,\sigma_p)$ and $u_n(\vec{k}_n, \sigma_n)$ are the
Dirac wave functions of the proton and neutron \cite{Ivanov2018};
$f_3(0)$ and $g_2(0)$ are the phenomenological coupling constants
defining the strength of the second class currents in the weak decays.
The contributions of the second class currents Eq.(\ref{eq:41}) to the
amplitude of the neutron $\beta^-$--decay in the non--relativistic
baryon approximation is defined by \cite{Ivanov2017d}
\begin{eqnarray}\label{eq:42}
 M(n \to p e^- \bar{\nu}_e)_{G - \rm odd} &=& -\,2m_n\,
\frac{G_F}{\sqrt{2}}\,V_{ud}\,\Big\{f_3(0)\,\frac{m_e}{M}\,
     [\varphi^{\dagger}_p\varphi_n] [\bar{u}_e (1 - \gamma^5)
       v_{\bar{\nu}}] +
     g_2(0)\,\frac{1}{M}\,[\varphi^{\dagger}_p(\vec{\sigma}\cdot
       \vec{k}_p)\varphi_n][\bar{u}_e \gamma^0 (1 - \gamma^5)
       v_{\bar{\nu}}]\nonumber\\
\hspace{-0.3in} && -
g_2(0)\,\frac{E_0}{M}\,[\varphi^{\dagger}_p\vec{\sigma}\varphi_n]
\cdot [\bar{u}_e \vec{\gamma}\,(1 - \gamma^5) v_{\bar{\nu}}] \Big\},
\end{eqnarray}
where we have kept only the leading $1/M$ terms in the large baryon
mass expansion.  The hermitian conjugate contribution is
\begin{eqnarray}\label{eq:43}
M^{\dagger}(n \to p e^- \bar{\nu}_e)_{G - \rm odd} &=&
-\,2m_n\,
\frac{G_F}{\sqrt{2}}\,V_{ud}\,\Big\{f^*_3(0)\,\frac{m_e}{M}\,
     [\varphi^{\dagger}_n\varphi_p] [\bar{v}_{\nu} (1 + \gamma^5) u_e]
     + g^*_2(0)\,\frac{1}{M}\,[\varphi^{\dagger}_n(\vec{\sigma}\cdot
       \vec{k}_p)\varphi_p][\bar{v}_{\nu}\gamma^0 (1 - \gamma^5)
       u_e]\nonumber\\ && -
     g^*_2(0)\,\frac{E_0}{M}\,[\varphi^{\dagger}_n\vec{\sigma}\varphi_p]
     \cdot [\bar{v}_{\nu} \vec{\gamma}\,(1 - \gamma^5) u_e] \Big\}.
\end{eqnarray}
The contributions of the $G$--odd correlations to the squared absolute
value of the amplitude of the neutron $\beta^-$--decay of polarized
electron and unpolarized neutron and proton, summed over polarizations
of massive fermions, are equal to
\begin{eqnarray}\label{eq:44}
 \hspace{-0.3in}&&\sum_{\rm pol.}\Big(M^{\dagger}(n \to p e^-
 \bar{\nu}_e)M (n \to p e^- \bar{\nu}_e)_{G - \rm odd} + M^{\dagger}
 (n \to p e^- \bar{\nu}_e)_{G - \rm odd}M(n \to p e^-
 \bar{\nu}_e)\Big) = 8m^2_n G^2_F|V_{ud}|^2 \nonumber\\
\hspace{-0.3in}&&\times \Big\{2\,{\rm
  Re}f_3(0)\,\frac{m_e}{M}\,\Big[\frac{m_e}{E_e} +
  \Big(\zeta^0_e\,\frac{\vec{k}_e\cdot \vec{k}_{\nu}}{E_e E_{\nu}} -
  \frac{\vec{\zeta}_e\cdot \vec{k}_{\nu}}{E_{\nu}}\Big)\Big] + 2\,{\rm
  Im}f_3(0)\,\frac{m_e}{M}\,\frac{\vec{\xi}_e\cdot (\vec{k}_e \times
  \vec{k}_{\nu})}{E_eE_{\nu}} + 2\,\lambda\,{\rm
  Re}g_2(0)\Big[\frac{1}{M}\Big(E_{\nu} +
  \frac{k^2_e}{E_e}\Big)\nonumber\\
\hspace{-0.3in}&&+ \frac{E_e + E_{\nu}}{M}\, \frac{\vec{k}_e\cdot
  \vec{k}_{\nu}}{E_e E_{\nu}} - \frac{E_e +
  E_{\nu}}{M}\,\frac{m_e}{E_e}\,\zeta^0_e -
\frac{m_e}{M}\,\zeta^0_e\,\frac{\vec{k}_e\cdot \vec{k}_{\nu}}{E_e E_{\nu}} -
\frac{m_e}{M}\, \frac{E_{\nu}}{E_e}\,\frac{\vec{\zeta}_e\cdot
  \vec{k}_{\nu}}{E_{\nu}} + \frac{E_0}{M}\Big(3 -
3\,\frac{m_e}{E_e}\,\zeta^0_e - \frac{\vec{k}_e\cdot
  \vec{k}_{\nu}}{E_e E_{\nu}} +
\frac{m_e}{E_e}\,\frac{\vec{\zeta}_e\cdot
  \vec{k}_{\nu}}{E_{\nu}}\Big)\Big]\nonumber\\
 \hspace{-0.3in}&&+ 2\,\lambda\,{\rm
  Im}g_2(0)\,\frac{m_e}{M}\, \frac{\vec{\xi}_e\cdot
  (\vec{k}_e \times \vec{k}_{\nu})}{E_eE_{\nu}}\Big\}.
\end{eqnarray}
For the relative $G$--odd contributions to the correlation
coefficients we obtain the following expressions
\begin{eqnarray}\label{eq:45}
 \hspace{-0.3in}\frac{\delta \zeta(E_e)_{G-\rm odd}}{\zeta^{(\rm
     SM)}(E_e)} &=& \frac{2}{1 + 3\lambda^2}\,\frac{1}{M}\,\Big\{{\rm
   Re}f_3(0)\,\frac{m^2_e}{E_e} + \lambda\,{\rm Re}g_2(0)\,\Big(4 E_0
 - \frac{m^2_e}{E_e}\Big)\Big\},\nonumber\\
\hspace{-0.3in}\frac{\delta a(E_e)_{G-\rm odd}}{a^{(\rm
     SM)}(E_e)} &=& - \delta \zeta(E_e)_{G-\rm odd},\nonumber\\
\hspace{-0.3in}\frac{\delta G(E_e)_{G-\rm odd}}{G^{(\rm SM)}(E_e)}
&=&\frac{2\lambda}{1 + 3\lambda^2}\,\frac{4E_0}{M}\, {\rm Re}g_2(0) -
\delta \zeta(E_e)_{G-\rm odd},\nonumber\\
\hspace{-0.3in}\frac{\delta H(E_e)_{G-\rm odd}}{H^{(\rm SM)}(E_e)}
&=&\frac{2}{1 - \lambda^2}\,\frac{E_e}{M}\,\Big({\rm Re}f_3(0) -
\lambda\,{\rm Re}g_2(0)\Big) - \delta \zeta(E_e)_{G-\rm odd},
\nonumber\\
\hspace{-0.3in}\frac{\delta K_e(E_e)_{G-\rm odd}}{K^{(\rm SM)}_e(E_e)}
&=&\frac{2}{1 - \lambda^2}\,\frac{m_e}{M}\,\Big(- {\rm Re}f_3(0) +
\lambda\,{\rm Re}g_2(0)\Big) - \delta \zeta(E_e)_{G-\rm odd},
\nonumber\\
\hspace{-0.3in}\frac{\delta L(E_e)_{G-\rm odd}}{L^{(\rm SM)}_e(E_e)}
&=& \frac{2}{1 - \lambda^2}\,\frac{k_e}{ \alpha M}\,\Big({\rm
  Im}f_3(0) + \lambda\,{\rm Im}g_2(0)\Big).
\end{eqnarray}
These expressions agree well with the G--odd correlations obtained in
\cite{Ivanov2017d} and as well as with those by Gardner and Plaster
\cite{Gardner2013}. For $\lambda = - 1.2750$ \cite{Abele2008} we get
\begin{eqnarray*}
 \hspace{-0.3in}\frac{\delta \zeta(E_e)_{G-\rm odd}}{\zeta^{(\rm
     SM)}(E_e)} &=& 1.85\times 10^{-4}\,{\rm
   Re}f_3(0)\,\frac{m_e}{E_e} + \Big(- 2.39\times 10^{-3} + 2.36\times
 10^{-4}\,\frac{m_e}{E_e}\Big)\,{\rm Re}g_2(0), \nonumber\\
\hspace{-0.3in}\frac{\delta a(E_e)_{G-\rm odd}}{a^{(\rm SM)}(E_e)} &=&
- 1.85\times 10^{-4}\,{\rm Re}f_3(0)\,\frac{m_e}{E_e} +
\Big(2.39\times 10^{-3} - 2.36\times
10^{-4}\,\frac{m_e}{E_e}\Big)\,{\rm Re}g_2(0), \nonumber\\
\hspace{-0.3in}\frac{\delta G(E_e)_{G-\rm odd}}{G^{(\rm SM)}(E_e)} &=&
- 1.85\times 10^{-4}\,{\rm Re}f_3(0)\,\frac{m_e}{E_e} - 2.36\times
10^{-4}\,{\rm Re}g_2(0)\,\frac{m_e}{E_e},\nonumber\\
\hspace{-0.3in}\frac{\delta H(E_e)_{G-\rm odd}}{H^{(\rm SM)}(E_e)} &=&
\Big(- 4.40\times 10^{-3}\frac{E_e}{E_0} - 1.85\times
10^{-4}\,\frac{m_e}{E_e}\Big)\,{\rm Re}f_3(0)\nonumber\\
\hspace{-0.3in}&& + \Big(2.39\times 10^{-3} - 5.61\times
10^{-3}\,\frac{E_e}{E_0} - 2.36\times 10^{-4}\,\frac{m_e}{E_e}
\Big)\,{\rm Re}g_2(0), \nonumber\\
\end{eqnarray*}
\begin{eqnarray}\label{eq:46}
\hspace{-0.3in}\frac{\delta K_e(E_e)_{G-\rm odd}}{K^{(\rm SM)}_e(E_e)}
&=&\Big(1.74\times 10^{-3} - 1.85\times
10^{-4}\,\frac{m_e}{E_e}\Big)\,{\rm Re}f_3(0) + \Big(4.61\times
10^{-3} - 2.36\times 10^{-4}\,\frac{m_e}{E_e}\Big)\,{\rm Re}g_2(0),
\nonumber\\
\hspace{-0.3in}\frac{\delta L(E_e)_{G-\rm odd}}{L^{(\rm SM)}(E_e)} &=&
\frac{k_e}{E_0}\,\Big(- 0.603\,{\rm Im}f_3(0) + 0.769\,{\rm
  Im}g_2(0)\Big).
\end{eqnarray}
Following Gardner and Plaster \cite{Gardner2013} and setting $f_3(0) =
0$ and $|{\rm Re}g_2(0)| < 0.01$ we obtain the contributions of the
G--odd correlations at the level of $10^{-5}$. Of course, the same
order of magnitude of the $G$--odd correlations one may get also for
$|{\rm Re}f_3(0)| < 0.01$ \cite{Ivanov2017d}.

\section{Discussion}
\label{sec:conclusion}

We have analysed the electron--energy and electron--antineutrino
angular distribution of the neutron $\beta^-$--decay with polarized
electron and unpolarized neutron and proton. The correlation
coefficients are calculated in the SM to order $10^{-3}$, caused by
the weak magnetism and proton recoil of order $O(E_e/M)$ and radiative
corrections of order $O(\alpha/\pi)$ Eqs.(\ref{eq:27}) -
(\ref{eq:29}). The radiative corrections to the correlation
coefficients $H(E_e)$ and $K_e(E_e)$ are defined by the functions
$(\alpha/\pi)\,h^{(3)}_n(E_e)$ and $(\alpha/\pi)\,h^{(4)}_n(E_e)$ (see
Eq.(\ref{eq:A.8}) in the Appendix), respectively, which have been never
calculated in literature. The correlation coefficients are also
supplemented by Wilkinson's higher order corrections Eq.(\ref{eq:30})
(see also Chapter III of Ref.\cite{Ivanov2017b}), which have not been
taken in Eqs.(\ref{eq:27}) - (\ref{eq:29}) and are induced by i) the
proton recoil in the Coulomb electron--proton final--state
interaction, ii) the finite proton radius, iii) the proton--lepton
convolution and iv) the higher--order {\it outer} radiative
corrections \cite{Wilkinson1982}.

Taking into account the contribution of interactions beyond the SM we
have arrived at the set of correlation coefficients $X_{\rm eff}(E_e)$
with $X = a,G,H$ and $K_e$, given in Eq.(\ref{eq:38}) and
Eq.(\ref{eq:39}). The structure of these contributions agrees well
with the results obtained in
\cite{Jackson1957}--\cite{Severijns2006}. These correlation
coefficients are presented in the form suitable for the analysis of
experimental data on searches of interactions beyond the SM at the
level of $10^{-4}$ \cite{Abele2016} (see also
\cite{Ivanov2013,Ivanov2017d}). The analysis of the supperallowed $0^+
\to 0^+$ transitions, carried out by Hardy and Towner \cite{Hardy2015}
and Gonz\'alez--Alonso {\it et al.} \cite{Severijns2018}, has shown
that in the approximation of real scalar coupling constants such as
$C_S = - \bar{C}_S$, i.e. the neutron and proton couple to
right--handed electron and antineutrino, the scalar coupling constants
are constrained by $|C_S| = 0.0014(13)$ and $|C_S| = 0.0014(12)$. Such
a small value of the scalar coupling constants commensurable with zero
can be justified by the property of the scalar density
$\bar{\psi}_p\psi_n $ with respect to the $G$--transformation
\cite{Lee1956a,Weinberg1958} (see also
\cite{Ivanov2017,Ivanov2018}). Indeed, the scalar density
$\bar{\psi}_p\psi_n = \bar{\psi}_N\tau^{(+)}\psi_N$, where $\psi_N$ is
the field operator of the nucleon isospin doublet with components
$(\psi_p, \psi_n)$ and $\tau^{(+)} = (\tau^1 + i\tau^2)/2$ is the
isospin $2\times 2$ Pauli matrix such as $\vec{\tau} = (\tau^1,
\tau^2,\tau^3)$ \cite{Itzykson1980}, is $G$--odd
\cite{Ivanov2017,Ivanov2018}. According to Weinberg
\cite{Weinberg1958}, the contributions of $G$--odd hadronic currents
or second class hadronic currents to the weak decays are suppressed
with respect to the contributions of $G$--even or first class hadronic
currents. As a result one may expect that in the neutron
$\beta^-$--decays the contributions of the tensor density
$\bar{\psi}_p\sigma_{\mu\nu}\psi_n = \bar{\psi}_N
\sigma_{\mu\nu}\tau^{(+)} \psi_N$, which is $G$--even
\cite{Ivanov2017,Ivanov2018}, should be larger than the contribution
of the scalar density $\bar{\psi}_p\psi_n =
\bar{\psi}_N\tau^{(+)}\psi_N$, which is $G$--odd
\cite{Ivanov2017,Ivanov2018}. These estimates agree well with the
contributions of order $10^{-5}$ of $G$--odd terms in the matrix
element of the hadronic $n \to p$ transition to the correlation
coefficients, which we have calculated in section \ref{sec:gparity} in
agreement with the results obtained by Gardner and Plaster
\cite{Gardner2013} and Ivanov {\it et al.} \cite{Ivanov2017d}.

It is obvious that the analysis of experimental data of experiments on
the searches of contributions of interactions beyond the SM at the
level of $10^{-4}$ or even better \cite{Abele2016} demands a robust SM
theoretical background with corrections at the level of
$10^{-5}$. These are i) Wilkinson's corrections \cite{Ivanov2017b} and
ii) corrections of order $O(E^2_e/M^2)$ defined by the weak magnetism
and proton recoil, calculated to next--to--next--to--leading order in
the large nucleon mass expansion, the radiative corrections of order
$O(\alpha E_e/M)$, calculated to next--to--leading order in the large
nucleon mass expansion, and the radiative corrections of order
$O(\alpha^2/\pi^2)$, calculated to leading order in the large nucleon
mass expansion \cite{Ivanov2017c}.  These theoretical corrections
should provide for the analysis of experimental data of "discovery"
experiments the required $5\sigma$ level of experimental uncertainties
of a few parts in $10^{-5}$ \cite{Ivanov2017b}. An important role of
strong low--energy interactions for a correct gauge invariant
calculation of radiative corrections of order $O(\alpha E_e/M)$ and
$O(\alpha^2/\pi^2)$ as functions of the electron energy $E_e$ has been
pointed out in \cite{Ivanov2017c}. This agrees with Weinberg's
assertion about important role of strong low--energy interactions in
decay processes \cite{Weinberg1957}. A procedure for the calculation
of these radiative corrections to the neutron $\beta^-$--decays with a
consistent account for contributions of strong low--energy
interactions, leading to gauge invariant observable expressions
dependent on the electron energy $E_e$ determined at the confidence
level of Sirlin's radiative corrections \cite{Sirlin1967}, has been
proposed in \cite{Ivanov2017c}. As we have shown that the
contributions of the $G$--odd correlations are at the level of
$10^{-5}$. Hence, the SM corrections of order $10^{-5}$ should be
important also as a theoretical background for the analysis of
experimental data on the search of the contributions of the $G$--odd
correlations in the neutron $\beta^-$--decays.

Finally, we would like to make some comments on the radiative
corrections of order $O(\alpha/\pi)$, which we have calculated to the
correlation coefficients of the electron--energy and
electron--antineutrino angular distribution of the neutron
$\beta^-$--decay with polarized electron and unpolarized neutron and
proton. Such a calculation has been performed in analogy to the
calculation of radiative corrections to the neutron lifetime and the
correlation coefficients of the electron--energy and
electron--antineutrino angular distributions of the neutron
$\beta^-$-decay with polarized neutron and unpolarized proton and
electron, carried out by Sirlin \cite{Sirlin1967} and Shann
\cite{Shann1971} (see also \cite{Gudkov2006} and \cite{Ivanov2013}),
and of the neutron $\beta^-$--decay with polarized neutron and
electron and unpolarized proton \cite{Ivanov2017b}. The radiative
corrections to the correlation coefficients define the level of
accuracy of the theoretical background for the measurements of these
correlation coefficients.  However, as has been pointed out by Gl\"uck
\cite{Gluck1998}, these results may not be applicable to precise
analyses of recoil measurements, dealing with recoil energy and
angular distributions. For the neutron $\beta^-$--decay with polarized
neutron and unpolarized proton and electron the radiative corrections
to the proton recoil--energy and angular distribution have been
calculated in \cite{Ivanov2013a} (see also \cite{Gudkov2006}). The
calculation of radiative corrections to the electron--proton
recoil--energy and angular distribution for the neutron
$\beta^-$--decay with polarized electron and unpolarized neutron and
proton demands a special consideration (see, for example,
\cite{Ivanov2013a}) and goes beyond the scope of this paper. We are
planning to perform such a calculation in our forthcoming publication.

\section{Acknowledgements}

We thank Hartmut Abele for stimulating discussions. The work of
A. N. Ivanov was supported by the Austrian ``Fonds zur F\"orderung der
Wissenschaftlichen Forschung'' (FWF) under contracts P31702-N27,
P26781-N20 and P26636-N20 and ``Deutsche F\"orderungsgemeinschaft''
(DFG) AB 128/5-2. The work of R. H\"ollwieser was supported by the
Deutsche Forschungsgemeinschaft in the SFB/TR 55. The work of
M. Wellenzohn was supported by the MA 23 (FH-Call 16) under the
project ``Photonik - Stiftungsprofessur f\"ur Lehre''.

\section*{Appendix A: Neutron radiative $\beta^-$--decay with polarized
 electron and unpolarized neutron, proton and photon}
\renewcommand{\theequation}{A-\arabic{equation}}
\setcounter{equation}{0}

In this Appendix we calculate the electron--energy and angular
distribution of the rate of the neutron radiative $\beta^-$--decay $n
\to p + e^- + \bar{\nu}_e + \gamma$ with polarized electron and
unpolarized neutron and proton, and photon. Such a distribution is
important for the cancellation of infrared divergences in the neutron
lifetime and correlation coefficients of the neutron $\beta^-$--decay
\cite{Sirlin1967,Ivanov2013,Ivanov2017b}. 

Following \cite{Ivanov2013,Ivanov2017b} (see also
\cite{Ivanov2013a,Ivanov2017c}) the energy and angular distribution of
the neutron radiative $\beta^-$--decay with polarized electron and
unpolarized neutron and proton is
\begin{eqnarray}\label{eq:A.1}
\hspace{-0.3in}&&\frac{d^8\lambda_{\beta^-_c\gamma}(E_e,\vec{k}_e,\vec{\xi}_e,
  \vec{k}_{\nu},\vec{q}\,)_{\lambda \lambda'}}{d\omega d E_e d\Omega_e
  d\Omega_{\nu} d\Omega_{\gamma}} = \frac{\alpha}{2\pi}\,(1 + 3
\lambda^2)\,\frac{G^2_F|V_{ud}|^2}{(2\pi)^6}\,\sqrt{E^2_e -
  m^2_e}\,F(E_e, Z = 1)\,\frac{(E_0 - E_e - \omega)^2}{(E_e -
  \vec{n}_{\vec{q}}\cdot \vec{k}_e)^2}\,\frac{1}{\omega}\nonumber\\
\hspace{-0.3in}&&\times \frac{1}{16} \Big\{{\rm tr}\{(\hat{k}_e + m_e
\gamma^5 \hat{\zeta}_e) Q_{\lambda} \gamma^0 \bar{Q}_{\lambda'}(1 -
\gamma^5)\} + a_0 \frac{\vec{k}_{\nu}}{E_{\nu}}\cdot {\rm
  tr}\{(\hat{k}_e + m_e \gamma^5 \hat{\zeta}_e) Q_{\lambda}
\vec{\gamma}\,\bar{Q}_{\lambda'}(1 - \gamma^5)\}\Big\},
\end{eqnarray}
where $d\Omega_e$, $d\Omega_{\nu}$ and $d\Omega_{\gamma}$ are elements
of the solid angels of the electron, antineutrino and photon,
respectively. Then, $Q_{\lambda} = 2 \varepsilon^*_{\lambda}(q) \cdot
k_e + \hat{\varepsilon}^*_{\lambda}(q)\,\hat{q}$ and
$\bar{Q}_{\lambda'} = \gamma^0 Q^{\dagger}_{\lambda'} \gamma^0 = 2
\varepsilon_{\lambda'}(q)\cdot k_e +
\hat{q}\,\hat{\varepsilon}_{\lambda'}(q) $, where
$\varepsilon^*_{\lambda}(q)$ (or $\varepsilon_{\lambda'}(q)$) and $q =
(\omega, \vec{q}\,) = (\omega, \omega \vec{n}_{\vec{q}})$ are the
polarization vector and 4--momentum of the photon obeying the
constraints $\varepsilon^*_{\lambda}(q)\cdot q = 0$ (or
$\varepsilon_{\lambda'}(q)\cdot q = 0$) and $q^2 = 0$,
$\vec{n}_{\vec{q}} = \vec{q}/\omega$ is a unit vector and
$\lambda(\lambda') = 1,2$ defines physical polarization states of the
photon. In Eq.(\ref{eq:A.1}) the traces over Dirac matrices in the
covariant form are defined by
\begin{eqnarray}\label{eq:A.2}
\hspace{-0.3in}&&\frac{1}{16}\,{\rm tr}\{\hat{a}\, Q_{\lambda}
\gamma^{\mu} \bar{Q}_{\lambda'}(1 - \gamma^5)\} =
(\varepsilon^*_{\lambda}\cdot k_e)(\varepsilon_{\lambda'}\cdot k_e)
a^{\mu} + \frac{1}{2}\Big( (\varepsilon^*_{\lambda}\cdot
k_e)(\varepsilon_{\lambda'}\cdot a) + (\varepsilon^*_{\lambda}\cdot
a)(\varepsilon_{\lambda'}\cdot k_e) - (\varepsilon^*_{\lambda}\cdot
\varepsilon^*_{\lambda})(a\cdot q)\Big) q^{\mu}\nonumber\\
\hspace{-0.3in}&& - \frac{1}{2}\Big((\varepsilon^*_{\lambda}\cdot k_e)
\varepsilon^{\mu}_{\lambda'} + \varepsilon^{*\mu}_{\lambda}
(\varepsilon_{\lambda'}\cdot k_e)\Big) (a\cdot q) -
\frac{1}{2}\,i\,\varepsilon^{\mu\nu\alpha\beta}\Big((\varepsilon^*_{\lambda}\cdot
k_e) \varepsilon_{\lambda' \nu} - \varepsilon^*_{\lambda
  \nu}(\varepsilon_{\lambda'}\cdot k_e)\Big) a_{\alpha} q_{\beta} -
\frac{1}{2}\,i\,q^{\mu} \varepsilon^{\rho\varphi\alpha\beta}
\varepsilon^*_{\lambda \rho} \varepsilon_{\lambda'
  \varphi}a_{\alpha}q_{\beta},~~
\end{eqnarray}
where $a = k_e$ and $a = m_e \zeta_e$, and
$\varepsilon^{\alpha\nu\mu\beta}$ is the Levi--Civita tensor defined
by $\varepsilon^{0123} = 1$ and $\varepsilon_{\alpha\nu\mu\beta}= -
\varepsilon^{\alpha\nu\mu\beta}$ \cite{Itzykson1980}. Plugging
Eq.(\ref{eq:A.2}) into Eq.(\ref{eq:A.1}), using the Coulomb gauge
\cite{Ivanov2013,Ivanov2017b} (see also
\cite{Ivanov2013a,Ivanov2017c}) and summing over photon polarizations
we obtain the following expression for the energy and angular
distribution of the neutron radiative $\beta^-$--decay
\begin{eqnarray}\label{eq:A.3}
\hspace{-0.15in}&&\frac{d^8\lambda_{\beta^-_c\gamma}(E_e,\vec{k}_e,\vec{\xi}_e,
  \vec{k}_{\nu},\vec{q}\,)}{d\omega d E_e d\Omega_e
  d\Omega_{\nu} d\Omega_{\gamma}} = \frac{\alpha}{2\pi}\,(1 + 3
\lambda^2)\,\frac{G^2_F|V_{ud}|^2}{(2\pi)^6}\,\sqrt{E^2_e -
  m^2_e}\,E_e\,F(E_e, Z = 1)\,(E_0 - E_e -
\omega)^2\,\frac{1}{\omega}\nonumber\\
\hspace{-0.15in}&&\times \Bigg\{\Big[\frac{\beta^2 -
    (\vec{n}_{\vec{q}}\cdot\vec{\beta})^2}{(1 -
    \vec{n}_{\vec{q}}\cdot\vec{\beta})^2}\Big(1 +
  \frac{\omega}{E_e}\Big) + \frac{1}{1 - \vec{n}_{\vec{q}}\cdot
    \vec{\beta}}\,\frac{\omega^2}{E^2_e}\Big] +
a_0\,\frac{\vec{k}_{\nu}}{E_{\nu}}\cdot \Big[\vec{\beta}\,
  \Big(\frac{\beta^2 - (\vec{n}_{\vec{q}}\cdot \vec{\beta}\,)^2}{(1 -
    \vec{n}_{\vec{q}}\cdot\vec{\beta}\,)^2} + \frac{1}{1 -
    \vec{n}_{\vec{q}}\cdot \vec{\beta}}\,\frac{\omega}{E_e}\Big)\nonumber\\
\hspace{-0.3in}&& + \vec{n}_{\vec{q}}\,\Big( -
\frac{m^2_e}{E^2_e}\,\frac{1}{(1 - \vec{n}_{\vec{q}}\cdot
  \vec{\beta}\,)^2}\,\frac{\omega}{E_e} + \frac{1}{1 -
  \vec{n}_{\vec{q}}\cdot \vec{\beta}}\,\frac{\omega}{E_e} + \frac{1}{1
  - \vec{n}_{\vec{q}}\cdot
  \vec{\beta}}\,\frac{\omega^2}{E^2_e}\Big)\Big] +
\frac{\vec{\xi_e}\cdot \vec{k}_e}{E_e}\Big\{\Big[ - \frac{\beta^2 -
    (\vec{n}_{\vec{q}}\cdot \vec{\beta}\,)^2}{(1 -
    \vec{n}_{\vec{q}}\cdot\vec{\beta}\,)^2}\nonumber\\
\hspace{-0.3in}&& - \frac{ 1}{(1 -
  \vec{n}_{\vec{q}}\cdot\vec{\beta}\,)^2}\,\frac{\omega}{E_e} -
\frac{1}{(1 -
  \vec{n}_{\vec{q}}\cdot\vec{\beta}\,)^2}\,\frac{\omega^2}{E^2_e}\Big]
+ \frac{m_e}{E_e}\Big[\Big( - \frac{
    \vec{n}_{\vec{q}}\cdot\vec{\zeta}_e}{1 -
    \vec{n}_{\vec{q}}\cdot\vec{\beta}}+ \frac{
    \vec{n}_{\vec{q}}\cdot\vec{\zeta}_e}{(1 -
    \vec{n}_{\vec{q}}\cdot\vec{\beta}\,)^2}\Big)\,\frac{\omega}{E_e} +
  \frac{ \vec{n}_{\vec{q}}\cdot\vec{\zeta}_e}{(1 -
    \vec{n}_{\vec{q}}\cdot\vec{\beta}\,)^2}\,\frac{\omega^2}{E^2_e}\Big]
\Big\} \nonumber\\
\hspace{-0.3in}&&-
a_0\,\frac{m_e}{E_e}\,\frac{\vec{k}_{\nu}}{E_{\nu}}\cdot
\Big\{\vec{\zeta}_e\, \frac{\beta^2 -
  (\vec{n}_{\vec{q}}\cdot\vec{\beta})^2}{(1 -
  \vec{n}_{\vec{q}}\cdot\vec{\beta})^2} + \vec{\beta}\,\frac{\zeta^0_e
  - \vec{n}_{\vec{q}}\cdot\vec{\zeta}_e}{(1 -
  \vec{n}_{\vec{q}}\cdot\vec{\beta})^2}\, \frac{\omega}{E_e} +
\vec{n}_{\vec{q}}\Big[\frac{\zeta^0_e}{1 -
    \vec{n}_{\vec{q}}\cdot\vec{\beta}}\, \frac{\omega}{E_e} +
  \frac{\zeta^0_e - \vec{n}_{\vec{q}}\cdot\vec{\zeta}_e}{(1 -
    \vec{n}_{\vec{q}}\cdot\vec{\beta})^2}\,
  \frac{\omega^2}{E^2_e}\Big]\Big\} \Bigg\}.
\end{eqnarray}
The integration over directions of the photon momentum we carry out
using the results obtain in the Appendix of Ref.\cite{Ivanov2017b}. As
result the energy and angular distribution Eq.(\ref{eq:A.3}) takes the
form
\begin{eqnarray}\label{eq:A.4}
\hspace{-0.15in}&&\frac{d^6\lambda_{\beta^-_c\gamma}(E_e,\vec{k}_e,\vec{\xi}_e,
  \vec{k}_{\nu},\vec{q}\,)}{d\omega d E_e d\Omega_e
  d\Omega_{\nu}} = \frac{\alpha}{\pi}\,(1 + 3
\lambda^2)\,\frac{G^2_F|V_{ud}|^2}{(2\pi)^5}\,\sqrt{E^2_e -
  m^2_e}\,E_e\,F(E_e, Z = 1)\,(E_0 - E_e -
\omega)^2\,\frac{1}{\omega}\nonumber\\
\hspace{-0.15in}&&\times \bigg\{\Big\{\Big(1 + \frac{\omega}{E_e} +
\frac{1}{2}\,\frac{\omega^2}{E^2_e}\Big)\Big[\frac{1}{\beta}\,{\ell
    n}\Big(\frac{1 + \beta}{1 - \beta}\Big) - 2\Big] +
\frac{\omega^2}{E^2_e}\Big\} + a_0\,\frac{\vec{k}_e\cdot
  \vec{k}_{\nu}}{E_eE_{\nu}}\,\Big[1 +
  \frac{1}{\beta^2}\,\frac{\omega}{E_e}\Big(1 +
  \frac{1}{2}\,\frac{\omega}{E_e}\Big)\Big]\Big[\frac{1}{\beta}\,{\ell
    n}\Big(\frac{1 + \beta}{1 - \beta}\Big) - 2\Big]\nonumber\\
\hspace{-0.3in}&& - \frac{\vec{\xi_e}\cdot \vec{k}_e}{E_e}\,\Big[1 +
  \frac{1}{\beta^2}\,\frac{\omega}{E_e}\Big(1 +
  \frac{1}{2}\,\frac{\omega}{E_e}\Big)\Big]
\Big[\frac{1}{\beta}\,{\ell n}\Big(\frac{1 + \beta}{1 - \beta}\Big) -
  2\Big] - \,\frac{\vec{\xi}_e\cdot \vec{k}_{\nu}}{E_{\nu}}\,
a_0\,\frac{m_e}{E_e}\,\Big(1 -
\frac{1}{2\beta^2}\,\frac{\omega^2}{E^2_e}\Big) \,
\Big[\frac{1}{\beta}{\ell n}\Big(\frac{1 + \beta}{1 - \beta}\Big) -
  2\Big]\nonumber\\
\hspace{-0.3in}&& - a_0\,\frac{(\vec{\xi}_e\cdot \vec{k}_e)(\vec{k}_e
  \cdot \vec{k}_{\nu})}{(E_e + m_e) E_e E_{\nu}} \Big\{\Big(1 -
\frac{1}{2\beta^2}\frac{\omega^2}{E^2_e}\Big)
\Big[\frac{1}{\beta}\,{\ell n}\Big(\frac{1 + \beta}{1 - \beta}\Big) -
  2\Big] + \big(1 + \sqrt{1 - \beta^2}\,
\big)\,\Big[\frac{1}{\beta^2}\,\frac{\omega}{E_e}\,
\Big[\frac{1}{\beta}\,{\ell n}\Big(\frac{1 + \beta}{1 - \beta}\Big) -
  2\Big]\nonumber\\
\hspace{-0.3in}&& + \frac{1}{2\beta^2}\,\frac{\omega^2}{E^2_e}
\,\Big(\frac{3 - \beta^2}{\beta^2}\Big[\frac{1}{\beta}\,{\ell
    n}\Big(\frac{1 + \beta}{1 - \beta}\Big) - 2\Big] -
2\Big)\Big]\Big\} \bigg\}.
\end{eqnarray}
The first three correlation coefficients agree well with the results,
obtained in \cite{Ivanov2013} (see Eq.(B-11) of Ref.\cite{Ivanov2013})
and \cite{Ivanov2017b} (see Eq.(A-5) of
Ref.\cite{Ivanov2017b}). Having integrated over the photon energy in
the region $\omega_{\rm min} \le \omega \le E_0 - E_e$, where
$\omega_{\rm in}$ is an infrared cut--off \cite{Ivanov2013}, we arrive
at the expression
\begin{eqnarray}\label{eq:A.5}
\hspace{-0.15in}&&\frac{d^5\lambda_{\beta^-_c\gamma}(E_e,\vec{k}_e,\vec{\xi}_e,
  \vec{k}_{\nu})}{d E_e d\Omega_e d\Omega_{\nu}} =
\frac{\alpha}{\pi}\,(1 + 3
\lambda^2)\,\frac{G^2_F|V_{ud}|^2}{(2\pi)^5}\,\sqrt{E^2_e -
  m^2_e}\,E_e\,F(E_e, Z = 1)\,(E_0 -
E_e)^2\Big\{g^{(1)}_{\beta^-_c\gamma}(E_e, \omega_{\rm
  min})\nonumber\\
\hspace{-0.15in}&&+ \frac{\vec{k}_e\cdot
  \vec{k}_{\nu}}{E_eE_{\nu}}\,a_0\,g^{(2)}_{\beta^-_c\gamma}(E_e,
\omega_{\rm min}) - \frac{\vec{\xi_e}\cdot
  \vec{k}_e}{E_e}\,g^{(2)}_{\beta^-_c\gamma}(E_e, \omega_{\rm min}) -
\frac{\vec{\xi}_e\cdot \vec{k}_{\nu}}{E_{\nu}}\,
a_0\,\frac{m_e}{E_e}\, g^{(5)}_{\beta^-_c\gamma}(E_e, \omega_{\rm
  min}) - a_0\,\frac{(\vec{\xi}_e\cdot \vec{k}_e)(\vec{k}_e \cdot
  \vec{k}_{\nu})}{(E_e + m_e) E_e E_{\nu}}\nonumber\\
\hspace{-0.15in}&&\times\, g^{(6)}_{\beta^-_c\gamma}(E_e, \omega_{\rm
  min})\Big\}.
\end{eqnarray}
The functions $g^{(1)}_{\beta^-_c\gamma}(E_e, \omega_{\rm min})$ and
$g^{(2)}_{\beta^-_c\gamma}(E_e, \omega_{\rm min})$ have been
calculated in \cite{Ivanov2013,Ivanov2017b}, whereas the functions
$g^{(5)}_{\beta^-_c\gamma}(E_e, \omega_{\rm min})$ and
$g^{(6)}_{\beta^-_c\gamma}(E_e, \omega_{\rm min})$ are defined by the
integrals
\begin{eqnarray}\label{eq:A.6}
\hspace{-0.15in}g^{(5)}_{\beta^-_c\gamma}(E_e, \omega_{\rm min}) &=&
\int^{E_0 - E_e}_{\omega_{\rm min}}\frac{d\omega}{\omega}\,\frac{(E_0
  - E_e - \omega)^2}{(E_0 - E_e)^2}\,\Big[\frac{1}{\beta}{\ell
    n}\Big(\frac{1 + \beta}{1 - \beta}\Big) - 2\Big]\,\Big(1 -
\frac{1}{2\beta^2}\,\frac{\omega^2}{E^2_e}\Big),\nonumber\\
\hspace{-0.15in}g^{(6)}_{\beta^-_c\gamma}(E_e, \omega_{\rm min}) &=&
\int^{E_0 - E_e}_{\omega_{\rm min}}\frac{d\omega}{\omega}\,\frac{(E_0
  - E_e - \omega)^2}{(E_0 -
  E_e)^2}\,\Big\{\Big[\frac{1}{\beta}\,{\ell n}\Big(\frac{1 +
    \beta}{1 - \beta}\Big) - 2\Big]\Big(1 -
\frac{1}{2\beta^2}\frac{\omega^2}{E^2_e}\Big)\nonumber\\
\hspace{-0.15in}&+& \big(1 + \sqrt{1 - \beta^2}\,
\big)\,\Big[\frac{1}{\beta^2}\,\frac{\omega}{E_e}\,
  \Big[\frac{1}{\beta}\,{\ell n}\Big(\frac{1 + \beta}{1 - \beta}\Big)
    - 2\Big] + \frac{1}{2\beta^2}\,\frac{\omega^2}{E^2_e}
  \,\Big(\frac{3 - \beta^2}{\beta^2}\Big[\frac{1}{\beta}\,{\ell
      n}\Big(\frac{1 + \beta}{1 - \beta}\Big) - 2\Big] -
  2\Big)\Big]\Big\}.\nonumber\\
\hspace{-0.15in}&&
\end{eqnarray}
The results of the integration are equal to
\begin{eqnarray}\label{eq:A.7}
\hspace{-0.15in}g^{(5)}_{\beta^-_c\gamma}(E_e, \omega_{\rm min}) &=&
\Big[{\ell n}\Big(\frac{E_0 - E_e}{\omega_{\rm min}}\Big) -
  \frac{3}{2} - \frac{1}{24 \beta^2}\,\frac{(E_0 - E_e)^2}{
    E^2_e}\Big]\,\Big[\frac{1}{\beta}{\ell n}\Big(\frac{1 + \beta}{1 -
    \beta}\Big) -
  2\Big],\nonumber\\ \hspace{-0.15in}g^{(6)}_{\beta^-_c\gamma}(E_e,
\omega_{\rm min}) &=& \Big[{\ell n}\Big(\frac{E_0 - E_e}{\omega_{\rm
      min}}\Big) - \frac{3}{2} - \frac{1}{24 \beta^2}\,\frac{(E_0 -
    E_e)^2}{ E^2_e}\Big]\,\Big[\frac{1}{\beta}{\ell n}\Big(\frac{1 +
    \beta}{1 - \beta}\Big) - 2\Big] \nonumber\\
\hspace{-0.15in}&+& \Big(1 + \sqrt{1 - \beta^2}\,
\Big)\,\frac{1}{3}\,\frac{(E_0 - E_e)}{\beta^2
  E_e}\Big\{\Big[\frac{1}{\beta}\,{\ell
    n}\Big(\frac{1 + \beta}{1 - \beta}\Big) - 2\Big] \nonumber\\
\hspace{-0.15in}&+& \frac{1}{8}\, \frac{E_0 - E_e}{E_e}\,\Big(\frac{3
  - \beta^2}{\beta^2}\Big[\frac{1}{\beta}\,{\ell n}\Big(\frac{1 +
    \beta}{1 - \beta}\Big) - 2\Big] - 2\Big)\Big]\Big\}
\end{eqnarray}
Now we are able to define the electron--energy and
electron--antineutrino angular distribution for the neutron
$\beta^-$--decay with polarized electron and unpolarized neutron and
proton, where the correlation coefficients are calculated to order
$10^{-3}$, caused by the weak magnetism and proton recoil of order
$O(E_e/M)$ and radiative corrections of order $O(\alpha/\pi)$.

The radiative corrections of order $O(\alpha/\pi)$ to the correlation
coefficients of the neutron $\beta^-$--decay with polarized electron
and unpolarized neutron and proton are defined by the function
$g_n(E_e)$ and the functions
\begin{eqnarray}\label{eq:A.8}
\hspace{-0.15in}f_n(E_e) &=&\lim_{\omega_{\rm min} \to
  0}[g^{(2)}_{\beta^-_c\gamma}(E_e,\omega_{\rm min}) -
  g^{(1)}_{\beta^-_c\gamma}(E_e,\omega_{\rm min})] +
g_F(E_e)\,\frac{m_e}{E_e} = \frac{1}{3}\,\frac{1 - \beta^2}{\beta^2}
\frac{E_0 - E_e}{E_e} \Big(1 + \frac{1}{8}\,\frac{E_0 -
  E_e}{E_e}\Big)\nonumber\\
\hspace{-0.15in}&&\times\,\Big[\frac{1}{\beta}\,{\ell n}\Big(\frac{1 +
    \beta}{1 - \beta}\Big) - 2\Big] - \frac{1}{12}\,\frac{(E_0 -
  E_e)^2}{E^2_e} + \frac{1 - \beta^2}{2 \beta}\,{\ell n}\Big(\frac{1 +
  \beta}{1 - \beta}\Big),\nonumber\\
\hspace{-0.15in}h^{(3)}_n(E_e) &=& \lim_{\omega_{\rm min} \to
  0}[g^{(5)}_{\beta^-_c\gamma}(E_e,\omega_{\rm min}) -
  g^{(1)}_{\beta^-_c\gamma}(E_e,\omega_{\rm min})] +
g_F(E_e)\,\frac{m_e}{E_e} - g_F(E_e)\,\frac{E_e}{m_e} = \nonumber\\
\hspace{-0.15in}&=& - \frac{1}{3}\,\frac{E_0 - E_e}{E_e}\Big\{ \Big(1 +
  \frac{1 + \beta^2}{8 \beta^2}\,\frac{E_0 -
    E_e}{E_e}\Big)\,\Big[\frac{1}{\beta}{\ell n}\Big(\frac{1 +
      \beta}{1 - \beta}\Big) - 2\Big] + \frac{1}{4}\,\frac{E_0 -
    E_e}{E_e}\Big\} - \frac{\beta}{2}\,{\ell n}\Big(\frac{1 + \beta}{1
  - \beta}\Big),\nonumber\\
\hspace{-0.15in}h^{(4)}_n(E_e) &=& \lim_{\omega_{\rm min} \to
  0}[g^{(6)}_{\beta^-_c\gamma}(E_e,\omega_{\rm min}) -
  g^{(1)}_{\beta^-_c\gamma}(E_e,\omega_{\rm min})] + g_F(E_e)\,
\frac{m_e}{E_e} + g_F(E_e) = \nonumber\\
\hspace{-0.15in}&=& - \frac{1}{3}\,\frac{E_0 - E_e}{E_e}\Big\{ \Big(1
+ \frac{1 + \beta^2}{8 \beta^2}\,\frac{E_0 -
  E_e}{E_e}\Big)\,\Big[\frac{1}{\beta}{\ell n}\Big(\frac{1 + \beta}{1
    - \beta}\Big) - 2\Big] + \frac{1}{4}\,\frac{E_0 - E_e}{E_e}\Big\}
+ \big(1 + \sqrt{1 - \beta^2}\,\big)
\nonumber\\
\hspace{-0.15in}&&\times \Big\{ \frac{1}{3}\,\frac{E_0 - E_e}{\beta^2
  E_e}\, \Big[\frac{1}{\beta}\,{\ell n}\Big(\frac{1 + \beta}{1 -
    \beta}\Big) - 2\Big] + \frac{1}{24}\,\frac{(E_0 - E_e)^2}{\beta^2
  E^2_e} \,\Big(\frac{3 - \beta^2}{\beta^2}\Big[\frac{1}{\beta}\,{\ell
    n}\Big(\frac{1 + \beta}{1 - \beta}\Big) - 2\Big] - 2\Big) \nonumber\\
\hspace{-0.15in}&& + \frac{\sqrt{1 - \beta^2}}{2 \beta}\, {\ell
  n}\Big(\frac{1 + \beta}{1 - \beta}\Big)\Big\}.
\end{eqnarray}
The functions $h^{(3)}_n(E_e)$ and $h^{(4)}_n(E_e)$ coincide with the
functions $h^{(1)}_n(E_e)$ and $h^{(2)}_n(E_e)$, calculated in
\cite{Ivanov2017b}.  For the calculation of the radiative corrections
to the neutron lifetime and correlation coefficients of the neutron
$\beta^-$--decay the integral
\begin{eqnarray}\label{eq:A.9}
\hspace{-0.3in}J(\beta, \kappa_{\,\rm IR}) =
\int\frac{d\omega}{\omega}\int\frac{d\Omega_{\gamma}}{4\pi}\,\frac{\beta^2
  - (\vec{n}_{\vec{q}}\cdot \vec{\beta}\,)^2}{(1 -
  \vec{n}_{\vec{q}}\cdot \vec{\beta}\,)^2},
\end{eqnarray}
which is logarithmically divergent in the infrared region of photon
energy \cite{Sirlin1967}, plays an important role. As has been pointed
out in \cite{Ivanov2013}, the result of the calculation of this
integral depends on the regularization procedure, where $\kappa_{\,\rm
  IR}$ is an infrared parameter.

Using the infrared cut--off regularization $\kappa_{\,\rm IR} =
\omega_{\rm min} \le \omega \le (E_0 - E_e)$, where $\omega_{\rm min}$
may be also treated as a photon--energy threshold of the detector, we
get
\begin{eqnarray}\label{eq:A.10}
\hspace{-0.3in}J(\beta, \omega_{\rm min}) = {\ell n}\Big(\frac{E_0 -
  E_e}{\omega_{\rm min}}\Big) \Big[\frac{1}{\beta}\,{\ell
    n}\Big(\frac{1 + \beta}{1 - \beta}\Big) - 2\Big].
\end{eqnarray}
In turn, the use of the finite photon--mass $\mu$ (FPM) regularization
\begin{eqnarray}\label{eq:A.11}
\hspace{-0.3in}J(\beta,\mu) = \int \frac{d^3q}{4\pi
  q^3_0}\,\frac{\beta^2 - (\vec{v}\cdot \vec{\beta}\,)^2}{(1 -
  \vec{v}\cdot \vec{\beta}\,)^2},
\end{eqnarray}
where $q_0 = \sqrt{\omega^2 + \mu^2}$ and $\vec{v} = \vec{q}/q_0$ are
energy and velocity of a photon with mass $\mu$, gives one (see
Eq.(B-26) of Ref.\cite{Ivanov2013})
\begin{eqnarray}\label{eq:A.12}
\hspace{-0.3in}J(\beta,\mu) = {\ell n}\Big(\frac{2(E_0 -
  E_e}{\mu}\Big)\Big[\frac{1}{\beta}\,{\ell n}\Big(\frac{1 + \beta}{1
    - \beta}\Big) - 2\Big] + 1 + \frac{1}{\beta}\,{\ell n}\Big(\frac{1
  + \beta}{1 - \beta}\Big) - \frac{1}{\beta}\,{\ell n}^2\Big(\frac{1 +
  \beta}{1 - \beta}\Big) - \frac{1}{2}\,{\rm Li}_2\Big(\frac{2\beta}{1
  + \beta}\Big),
\end{eqnarray}
where ${\rm Li}_2(x)$ is a Polylogarithmic function
\cite{Mitchel1949,Lewin1981}. We would like to emphasize that the
infinitesimal photon mass $\mu$, providing a Lorentz covariant
regularization of infrared divergences in the neutron
$\beta^-$--decays, cannot be identified with the infrared cut--off
$\omega_{\rm min}$, which can be treated as a photon--energy threshold
of the detector \cite{Nico2006,Cooper2010,Bales2016} (see also
\cite{Ivanov2013}). Nevertheless, the use of the Lorentz covariant FPM
regularization is important only for the calculation of the function
$g_n(E_e)$, defining the radiative corrections to the neutron lifetime
\cite{Sirlin1967}. It is required by gauge invariance of radiative
corrections and by the Kinoshita--Lee--Nauenberg theorem
\cite{Sirlin1967} (see also \cite{Ivanov2013}). In turn, for the
calculation of the functions $f_n(E_e)$ and $h^{(\ell)}_n(E_e)$, where
$\ell = 1,2$ \cite{Ivanov2017b} and $\ell = 3,4$ (see
Eq.(\ref{eq:A.8})), one may use both the Lorentz covariant FPM
regularization with an infinitesimal photon mass $\mu$ and the
infrared cut--off $\omega_{\rm min}$ regularization. Indeed, the
contributions of the integral $J(\beta, \kappa_{\rm IR})$, the
regularization of which depends on the regularization procedure (see
Eq.(\ref{eq:A.10}) and Eq.(\ref{eq:A.12})), cancel themselves in the
differences $\lim_{\kappa_{\rm IR} \to 0}
[g^{(i)}_{\beta^-_c\gamma}(E_e, \kappa_{\rm IR}) -
  g^{(1)}_{\beta^-_c\gamma}(E_e, \kappa_{\rm IR})]$, where $i =
2,3,4,5,6$, and the results do not depend on the regularization
procedure.

\end{document}